\makeatletter
\declare@file@substitution{revtex4-1.cls}{revtex4-2.cls}
\makeatother
\documentclass[twocolumn]{aastex631}
\usepackage{newtxtext,newtxmath}
\usepackage[T1]{fontenc}
\usepackage{ae,aecompl}
\usepackage{graphicx}
\usepackage{amsmath}
\usepackage{amssymb}	
\usepackage{float}
\usepackage[]{hyperref}

\usepackage{booktabs}
\accepted{for publication in PASA}
\usepackage{multirow}

\begin{document}

\title{Unravelling the role of merger histories in the population of In~situ stars: linking IllustrisTNG cosmological simulation to H3 survey }

\shorttitle{In~situ stars}
\shortauthors{R. Emami et. al.}

\correspondingauthor{Razieh Emami}
\email{razieh.emami$_{-}$meibody@cfa.harvard.edu}

\author[0000-0002-2791-5011]{Razieh Emami}
\affiliation{Center for Astrophysics $\vert$ Harvard \& Smithsonian, 60 Garden Street, Cambridge, MA 02138, USA}

\author[0000-0001-6950-1629]{Lars \ Hernquist}
\affiliation{Center for Astrophysics $\vert$ Harvard \& Smithsonian, 60 Garden Street,  Cambridge, MA 02138, USA}

\author[0000-0003-4284-4167]{Randall \ Smith}
\affiliation{Center for Astrophysics $\vert$ Harvard \& Smithsonian, 60 Garden Street, Cambridge, MA 02138, USA}

\author[0000-0002-5872-6061]{James F. Steiner}
\affiliation{Center for Astrophysics $\vert$ Harvard \& Smithsonian, 60 Garden Street, Cambridge, MA 02138, USA}

\author[0000-0001-6950-1629]{Grant Tremblay}
\affiliation{Center for Astrophysics $\vert$ Harvard \& Smithsonian, 60 Garden Street,  Cambridge, MA 02138, USA}

\author[0000-0003-2808-275X]{Douglas Finkbeiner}
\affiliation{Center for Astrophysics $\vert$ Harvard \& Smithsonian, 60 Garden Street, Cambridge, MA 02138, USA}

\author[0000-0001-8593-7692]{Mark \ Vogelsberger}
\affiliation{Department of Physics, Kavli Institute for Astrophysics and Space Research, Massachusetts Institute of Technology, Cambridge, MA 02139, USA}

\author[0000-0002-1323-5314]{Josh Grindlay}
\affiliation{Center for Astrophysics $\vert$ Harvard \& Smithsonian, 60 Garden Street, Cambridge, MA 02138, USA}

\author[0000-0003-3816-7028]{Federico Marinacci}
\affiliation{Department of Physics \& Astronomy ``Augusto Righi'', University of Bologna, via Gobetti 93/2, 40129 Bologna, Italy}
\affiliation{INAF, Astrophysics and Space Science Observatory Bologna, Via P. Gobetti 93/3, I-40129 Bologna, Italy}

\author[0000-0003-1598-0083]{Kung-Yi \ Su}
\affiliation{Black Hole Initiative at Harvard University, 20 Garden Street, Cambridge, MA 02138, USA}

\author[0000-0002-8791-6286]{Cecilia Garraffo}
\affiliation{Center for Astrophysics $\vert$ Harvard \& Smithsonian, 60 Garden Street, Cambridge, MA 02138, USA}

\author[0000-0001-5082-9536]{Yuan-Sen Ting}
\affiliation{Center for Cosmology and AstroParticle Physics (CCAPP), The Ohio State University, Columbus, OH 43210, USA}
\affiliation{Department of Astronomy, The Ohio State University, Columbus, USA}
\affiliation{Research School of Astronomy \& Astrophysics, Australian National University, Cotter Rd., Weston, ACT 2611, Australia}
\affiliation{Research School of Computer Science, Australian National University, Acton ACT 2601, Australia}

\author[0000-0002-1617-8917]{Phillip A. Cargile}
\affiliation{Center for Astrophysics $\vert$ Harvard \& Smithsonian, 60 Garden Street, Cambridge, MA 02138, USA}

\author[0000-0002-3324-4824]{Rebecca L. Davies}
\affiliation{Centre for Astrophysics and Supercomputing, Swinburne University of Technology, Hawthorn, Victoria, Australia} 
\affiliation{ARC Centre of Excellence for All Sky Astrophysics in 3 Dimensions (ASTRO 3D), Australia}

\author[0000-0001-5378-9998]{Chloë E. Benton}
\affiliation{Department for Astrophysical and Planetary Science, University of Colorado, Boulder, CO 80309, USA}

\author[0000-0002-0682-3310]{Yijia Li}
\affiliation{Department of Astronomy \& Astrophysics, The Pennsylvania State University, University Park, PA 16802, USA}
\affiliation{Institute for Gravitation and the Cosmos, The Pennsylvania State University, University Park, PA 16802, USA}

\author{Letizia Bugiani}
\affiliation{Dipartimento di Fisica e Astronomia, Università di Bologna, Bologna, Italy}

\author[0009-0009-6563-282X]{Amir H. Khoram}
\affiliation{Department of Physics \& Astronomy ``Augusto Righi'', University of Bologna, via Gobetti 93/2, 40129 Bologna, Italy}
\affiliation{INAF, Astrophysics and Space Science Observatory Bologna, Via P. Gobetti 93/3, I-40129 Bologna, Italy}

\author[0000-0002-0974-5266]{Sownak Bose}
\affiliation{Institute for Computational Cosmology, Department of Physics, Durham University, South Road, Durham DH1 3LE, UK}


\begin{abstract}
We undertake a comprehensive investigation into the distribution of in~situ stars within Milky Way-like galaxies, leveraging TNG50 simulations and comparing their predictions with data from the H3 survey. Our analysis reveals that 28\% of galaxies demonstrate reasonable agreement with H3, while only 12\% exhibit excellent alignment in their profiles, regardless of the specific spatial cut employed to define in~situ stars. To uncover the underlying factors contributing to deviations between TNG50 and H3 distributions, we scrutinise correlation coefficients among internal drivers(e.g., virial radius, star formation rate [SFR]) and merger-related parameters (such as the effective mass-ratio, mean distance, average redshift, total number of mergers, average spin-ratio and maximum spin alignment between merging galaxies). Notably, we identify significant correlations between deviations from observational data and key parameters such as the median slope of virial radius, mean SFR values, and the rate of SFR change across different redshift scans. Furthermore, positive correlations emerge between deviations from observational data and parameters related to galaxy mergers. We validate these correlations using the Random Forest Regression method. Our findings underscore the invaluable insights provided by the H3 survey in unravelling the cosmic history of galaxies akin to the Milky Way, thereby advancing our understanding of galactic evolution and shedding light on the formation and evolution of Milky Way-like galaxies in cosmological simulations.
\end{abstract}

\keywords{Milky Way Galaxy, H3 survey, Stellar distribution}

\section{Introduction}
The stellar distribution within the Milky Way (MW) galaxy offers valuable insights into its merger history. By combining photometric observations with stellar kinematics, we can reconstruct the accretion history of the MW. Within the MW, stellar components may originate from its main galaxy, including those formed from accreted gas, known as in~situ stars, or they may be accreted from satellite galaxies, termed ex-situ stars. Understanding whether the MW is predominantly composed of in~situ stars or enriched by ex-situ components provides a crucial avenue for advancing theories of galaxy formation and evolution.

The origin of the Milky Way's stellar halo has been a focus of theoretical investigation in several studies.  Utilising Sloan Digital Sky Survey (SDSS) Data Release 5 (DR5), \cite{2008ApJ...680..295B} observed that the structure of the MW's stellar halo resembles debris from a disrupted satellite galaxy, suggesting a substantial contribution from tidally disrupted galaxies. This conclusion was supported by \cite{2019MNRAS.482.3426M}, who employed abundance data from APOGEE-DR14 and kinematic data from Gaia-DR. Additionally, \cite{2021MNRAS.504.1657K} analysed APOGEE data, focusing on the chemical composition of nitrogen-rich stars, and estimated that approximately 30\% of the inner galactic stars originated from disrupted globular clusters (GCs). Furthermore, \cite{2019ApJ...874L..35M} identified a population of retrograde, metal-poor stars potentially originating from accreted dwarf galaxies, contributing to the understanding of the MW's stellar halo formation.

Numerous studies based on numerical simulations have investigated the fraction of in~situ stars in the Milky Way galaxy \cite[see, for example,][and references therein]{2009ApJ...702.1058Z, 2010MNRAS.404.1711P, 2015MNRAS.454.3185C, 2015ApJ...799..184P, 2019MNRAS.485.2589M, 2020MNRAS.497.4459F}. These analyses have yielded a wide range of inferred values for the in~situ component, with simulations suggesting it can vary greatly, from being relatively negligible to nearly comparable to the accreted stellar components.

Observationally, recent advancements in stellar spectroscopic surveys, including RAVE \citep{2006AJ....132.1645S}, SEGUE \citep{2009AJ....137.4377Y}, LAMOST \citep{2012RAA....12.1197C}, GALAH \citep{2015MNRAS.449.2604D}, APOGEE \citep{2017AJ....154...94M}, Gaia \citep{2016A&A...595A...1G}, and the Hectochelle in Halo at the High Resolution survey, hereafter the H3 survey, \citep{2019ApJ...883..107C, 2019ApJ...887..237C}, have provided precise information regarding the position, velocity, and chemical abundances of millions of stars in the solar neighbourhood.

Building upon insights from the H3 survey, \cite{2020ApJ...901...48N} demonstrated that the MW is predominantly composed of substructures that have been accreted onto the galaxy. 
They combined data from the H3 Survey with Gaia to construct a comprehensive six-dimensional phase-space, incorporating stellar chemical information, to reconstruct the stellar structure of the Milky Way. Their analysis focused on a sample of 5684 giant stars at $\vert b \vert > 40^{\circ} $ and $\vert Z \vert > 2$ kpc, within 50 kpc of the Galactic centre In addition to identifying previously known structures in the Milky Way, they uncovered several new stellar substructures. Notably, their findings revealed that beyond $\vert Z \vert > 15$ kpc, more than 80\% of the halo is composed of stars accreted from dwarf galaxies, providing crucial insights into the origins and assembly history of the Milky Way's stellar halo.

They specifically analysed the chemical abundance of stars spanning from the local halo to the extended stellar halo, shedding light on their origins.

Expanding on these findings regarding the origins of individual stars, the question arises concerning the origin and nature of the stellar halo. Specifically, there is a debate over whether the halo consists mainly of in~situ or ex-situ stars \citep[see, for example,][and references therein]{1962ApJ...136..748E, 1978ApJ...225..357S, 2021MNRAS.506.5410I, 2021ApJ...908..191C, 2021A&ARv..29....5M}. Additionally, understanding the radial extent of in~situ and ex-situ stars, along with their relative ratios in the MW galaxy, provides insights into the MW's accretion history. Recent studies, such as \cite{2022arXiv220804327H}, have shown that the stellar halo in the MW is tilted. The presence of a tilt in the stellar halo serves as a key indicator of an accreted stellar halo in the Milky Way and provides valuable insights into the dynamical evolution of past mergers. Furthermore, this tilt may reflect an underlying asymmetry in the dark matter (DM) distribution, with potential implications for galaxy evolution modelling and direct DM detection experiments.

In this study, we conduct an in-depth analysis of in~situ stars in 25 Milky Way-like galaxies simulated using the TNG50 run of the IllustrisTNG simulation \citep{2019MNRAS.490.3196P, 2019MNRAS.490.3234N}. We quantify the scale-height distribution of the in~situ star fraction and compare it with the distribution of in~situ stars from the H3 survey, \cite{2020ApJ...901...48N}. 
We employ various spatial cuts in defining the in~situ stars and compare the spatial distribution of stars from our galaxy sample with the ones from H3 survey. By visual inspection, we find that in 28\% of galaxies in our sample the spatial distribution of stars exhibit reasonable agreement with the observational data, while only in 12\% we see an 
excellent alignment between the stellar distribution from the theory and observation. 

We investigate the key drivers contributing to the discrepancy between the TNG50 results and the H3 Survey findings regarding the scale height dependence of the in-situ star fraction,  categorising the key drivers into internal and external factors.  Internal drivers pertain to the intrinsic properties of a given halo, while external drivers are relevant only in the context of galaxy mergers.  We infer the correlation of the deviation with both of internal factors, such as the virial radius and star formation rate) as well as external parameters related to mergers (including the merger mass ratio, mean merging galaxy distance, effective merger redshift, total number of mergers, mean spin fraction, and maximum alignment of merging galaxy spins.

Our findings reveal significant correlations between these parameters and the deviation from H3 observations, underscoring the utility of H3 results in providing valuable constraints on galaxy evolution.

The paper is organised as follows. Section \ref{TNG-H3} presents an overview of the TNG simulation as well as the H3 survey. Section \ref{Insitu}, introduces the in~situ stars from the TNG50 simulation, and Section \ref{Difference-TNG-H3}, explores the origin of the discrepancies between the simulation and observations. Section \ref{conclusion} presents the conclusion of the paper. Several technical details are presented in Appendices \ref{mass-ratio-eff} and \ref{Random-Forest-Regression}. 

\section{Sample selection in TNG vs the selection functions in H3}
\label{TNG-H3}

In this section, we provide a summary of the sample selections made from both the TNG50 simulation and the H3 survey. These selections form the foundation of the analysis presented throughout the remainder of the paper. 

\subsection{Milky Way-like galaxies in TNG50}
\label{TNG50-Sim}
The IllustrisTNG simulations represent the next generation of cosmological hydrodynamical simulations designed to model galaxy formation and evolution within the framework of the $\Lambda$CDM paradigm \citep{2019MNRAS.490.3196P, 2019MNRAS.490.3234N}. Building upon the foundations laid by the earlier Illustris simulations \citep{2014MNRAS.444.1518V, 2014Natur.509..177V, 2014MNRAS.445..175G, 2015MNRAS.452..575S}, IllustrisTNG incorporates significant improvements, particularly in the modelling of AGN feedback, chemical enrichment, and the evolution of seed magnetic fields \citep{2017MNRAS.465.3291W, 2018MNRAS.473.4077P}.

In this paper, we concentrate on the TNG50 simulation that operates within a periodic box with a size of L$_{\mathrm{box}}$ = 35 Mpc/h, containing 2160$^3$ gas elements and dark matter particles with mass resolutions of [0.85, 4.5] $\times 10^5 M_{\odot}$, respectively. Moreover, TNG50 adopts the cosmological parameters specified by \cite{2016A&A...594A..13P}. This simulation provides a detailed and comprehensive framework for studying the formation and evolution of galaxies across cosmic time.

We examine a carefully selected sample of Milky Way-like galaxies from TNG50 identified and characterised in previous works \citep{2020arXiv200909220E, 2020arXiv201212284E,2022ApJ...937...20E,2024ApJ...961..193W}. Our sample selection adheres to two primary criteria.  First, we constrain the dark matter halo mass to fall within the range of (1-1.6)$\times 10^{12} M_{\odot}$, motivated by recent observational constraints on the dark matter halo mass of the Milky Way galaxy \citep{2019A&A...621A..56P}.  Second, we limit our galaxy sample to rotationally supported galaxies identified in two steps, as the following. 

As a first step, we compute the stellar net specific angular momentum vector, $\mathbf{j}_{\rm{net}}$, for each galaxy in our sample:
\begin{equation}
\label{net-J}
\mathbf{j}_{\rm{net}} \equiv \frac{\mathbf{J}_{\rm{tot}}}{M} = \frac{\sum_{i} m_i \mathbf{r}_i \times \mathbf{v}_i}{\sum_{i} m_i},
\end{equation}
where the index $i$ refers to stellar particles. 
Aligning the $z$-axis with the direction of $\mathbf{j}_{\rm{net}}$, we compute the inner product of each stellar particle's angular momentum with the $z$-axis, given by $j_{z,i} = \mathbf{j}_i \cdot \mathbf{\hat{z}}$. The orbital circularity parameter is then defined as: 
\begin{equation}
\label{orbital-circularity}
\varepsilon_i \equiv \frac{j_{z,i}}{j_c(E_i)} ~~~,~~~ j_c(E_i) = r_c v_c = \sqrt{G M(\leq r_c) r_c}. 
\end{equation}
For each particle, we determine the radius of its corresponding circular orbit by equating its total energy to the specific energy of a circular orbit: 

\begin{equation}
E(r_c) = \frac{G M(\leq r_c)}{2r_c} + \phi(r_c),
\end{equation}
where $ M(\leq r_c) $ denotes the mass enclosed within the circular orbit, and $ \phi(r_c) $ represents the gravitational potential at $ r_c $,  computed from the averaged radial profile of the total gravitational potential, accounting for contributions from stars, gas, dark matter, and the central black hole. 

We define disk stars as those with $ \varepsilon_i \geq 0.7 $. Furthermore, we restrict our sample to cases where at least 40\% of the stars within a radial distance of 10 kpc from the centre exhibit disk-like properties. This criterion reduces our sample to 25 Milky Way-like galaxies.

\subsection{H3 observation and selection functions}
\label{H3-survey}
The H3 (Hectochelle in Halo at the High Resolution) Survey is an ongoing stellar spectroscopic survey that offers an unbiased measurement of stellar parameters \citep{2019ApJ...883..107C, 2019ApJ...887..237C}. It provides spectro-photometric distances for approximately 2$\times 10^5$ stars within the photometric magnitude range of $15 < r < 18$, with a 3D heliocentric distance of d$_{\mathrm{helio}} >$ 3 kpc, $|b| > 40 ^{\circ}$, and Dec $> -20^{\circ}$. H3 survey outputs radial velocities, spectroscopic distances, [Fe/H], and [$\alpha$/Fe] abundances for the aforementioned stellar sample. By combining this data with the Gaia proper motion measurements, enables the determination of the full 6D phase-space information and 2D chemical-space information for these stars.

In a related study, \cite{2020ApJ...901...48N} focused on 5684 K giants from the H3 survey and conducted an extensive exploration of the structure of distant galaxies up to 50 kpc from the galactic centre. They analysed the scale-height dependence of the in~situ stars. \cite{2020ApJ...901...48N}  used a dedicated approach, combining the kinematics information as well as the metallicity and made a general class of in~situ stars including the contributions from different parts such as the high-$\alpha$ disk, in~situ halo, the metal-weak thick disk, Aleph (defined mainly based on their specific values of [Fe/H] $>$ -0.8 and $[\alpha$/Fe] $<$ 0.27 together with some kinematic criteria such as the azimuthal component ($-300 km s^{-1} < V_{\phi} < -175 km s^{-1} $) and the radial component ( $ \vert V_r \vert < 75 km s^{-1}$) of the stellar velocity), and unclassified disk debris.

Motivated by these observational findings, our study investigates in~situ stars within a selected sample of Milky Way-like galaxies from the TNG50 simulation. We focus on assessing the scale height dependence of the in~situ star fraction and contrasting our results with those from the H3 survey. While the H3 survey employed a combination of the kinematics and chemical cuts involving [Fe/H] and [$\alpha$/Fe] abundances, alongside photometric magnitude, in this first study, our approach to identifying in~situ stars within the TNG50 simulations is based on the individual stellar birth locations being located within a given radial cut throughout the entire of the cosmic evolution.
Following the notation of \cite{2007MNRAS.375....2D}, the first progenitor is the one with the most massive assembly history of the main progenitor of the main subhalo in SubFind convention while the second progenitor is defined as the galaxy with
the second massive history behind it. In this paper, we often refer to the first ($f$) and second progenitor ($s$). Furthermore, to facilitate comparison with the H3 findings, we add  similar spatial selection functions and study the role of varying radial cutoffs to compile a sample of in~situ stars in our galaxy dataset. Subsequently, we conduct a comparative analysis of the scale height profile of in~situ stars with the H3 survey results, representing their distance from the galactic disk plane.

\section{Theoretical estimation of in~situ stars} \label{Insitu}

\subsection{Inferring the in~situ stars from TNG50 simulation}
\label{Def:insitu-TNG}
The in~situ stellar component comprises stars born within the disk of their galactic host halo, as well as stars formed from streams of stripped gas originating from infalling satellites \citep{2004MNRAS.351.1215B, 2009ApJ...702.1058Z, 2010MNRAS.404.1711P, 2012MNRAS.420.2245M, 2014MNRAS.439.3128T, 2015ApJ...799..184P, 2016MNRAS.458.2371R, 2019MNRAS.485.2589M}. In~situ stars born in the disk may be ejected to larger distances due various interactions. The amount of ejected mass may depend to some degree on subgrid physics and stellar feedback \citep[see, for instance,][and references therein]{2009ApJ...702.1058Z, 2015MNRAS.454.3185C}. Consequently, the radial profile of the fraction of in~situ stars varies among different simulations, ranging from negligible at larger distances ($>5$ kpc) \citep{2009ApJ...702.1058Z,2015ApJ...799..184P} to dominant at larger radii \citep{2011MNRAS.416.2802F, 2016MNRAS.459L..46M, 2016MNRAS.457.1419M, 2018MNRAS.479.4004E, 2019MNRAS.485.2589M}.
On the contrary, accreted stars are born in satellite galaxies, with the satellite either bound to the main progenitor from the outset or accreted into the main progenitor at a later time \citep[see, for example,][and references therein]{2014MNRAS.439.3128T,2019MNRAS.485.2589M}.

In the following, we present our strategy for selecting the in~situ stars for a sample of 25 MW-like galaxies in the TNG50 simulation. The in~situ star fraction is defined as the ratio of in~situ stars to the total number of stars at that location. For more details about our sample selection, please refer to Section \ref{TNG50-Sim}.

We use merger trees constructed from the sublink algorithm \citep{2015MNRAS.449...49R, 2016MNRAS.458.2371R} to identify the main progenitors of the subhalo, which are defined as the most massive progenitor of the target subhalo at any given cosmic time \citep{2007MNRAS.375....2D}. Furthermore, we employ baryonic merger trees, tracking only the stellar components in our analysis. As previously mentioned, we exclusively consider stars belonging to the main progenitor. Specifically, we initiate with the stellar components bound to the main halo at zero redshift, $z = 0$, and trace their birth locations backward in time. Subsequently, we ascertain whether each star was part of the main progenitor at its birth time. Finally, we determine the distance of the star from its host galaxy at its birth time and compare it with a radial threshold, denoted as $r_{\mathrm{cut}}$. We employ four different selections for $r_{\mathrm{cut}}$, encompassing either a constant value of $30$ kpc 
comoving radius or three time-dependent distances defined as some fixed coefficient of the virial distance (hereafter $R_{\mathrm{vir}}$) at the star's birth time; further details are provided below. If the star is situated within the specified threshold, we categorise it as an in~situ star; otherwise, it is considered as being accreted onto the main halo.

\subsection{Redshift evolution of the main progenitor}
\label{merger-history}
As previously mentioned, we use the  SubFind main subhalos generated by the sublink algorithm to investigate the halo merger history for the first and second progenitors of all halos in our MW-like sample. Through this process, we determine the mass-ratio and merger time for each merger event. Following the standard methodology outlined in \cite{2015MNRAS.449...49R, 2016MNRAS.458.2371R}, the mass-ratio in a merger system is determined when the second progenitor reaches its maximum mass. However, estimating the precise merger time is more intricate, as the second progenitor may orbit around the first progenitor and subsequently lose a significant portion of its mass. In our analysis below, we compute all correlation coefficients at the point when the second progenitor achieves its maximum mass.

When analysing the influence of galaxy mergers on variables such as the star formation rate (SFR) and the comoving virial radius ($R_{\mathrm{vir}}$), defined at the density threshold of 200 times the critical density, we categorise mergers into minor and major categories. Minor mergers are defined as those with a mass-ratio between 0.05 and 0.2, while major mergers are those with a mass-ratio above 0.2. However, for the computation of correlation coefficients, as described below, we focus exclusively on two scenarios: one involving mergers with a mass-ratio above 0.05 and the other involving mergers with a mass-ratio above 0.2.
\begin{figure*}
\center
\includegraphics[width=1.0\textwidth]{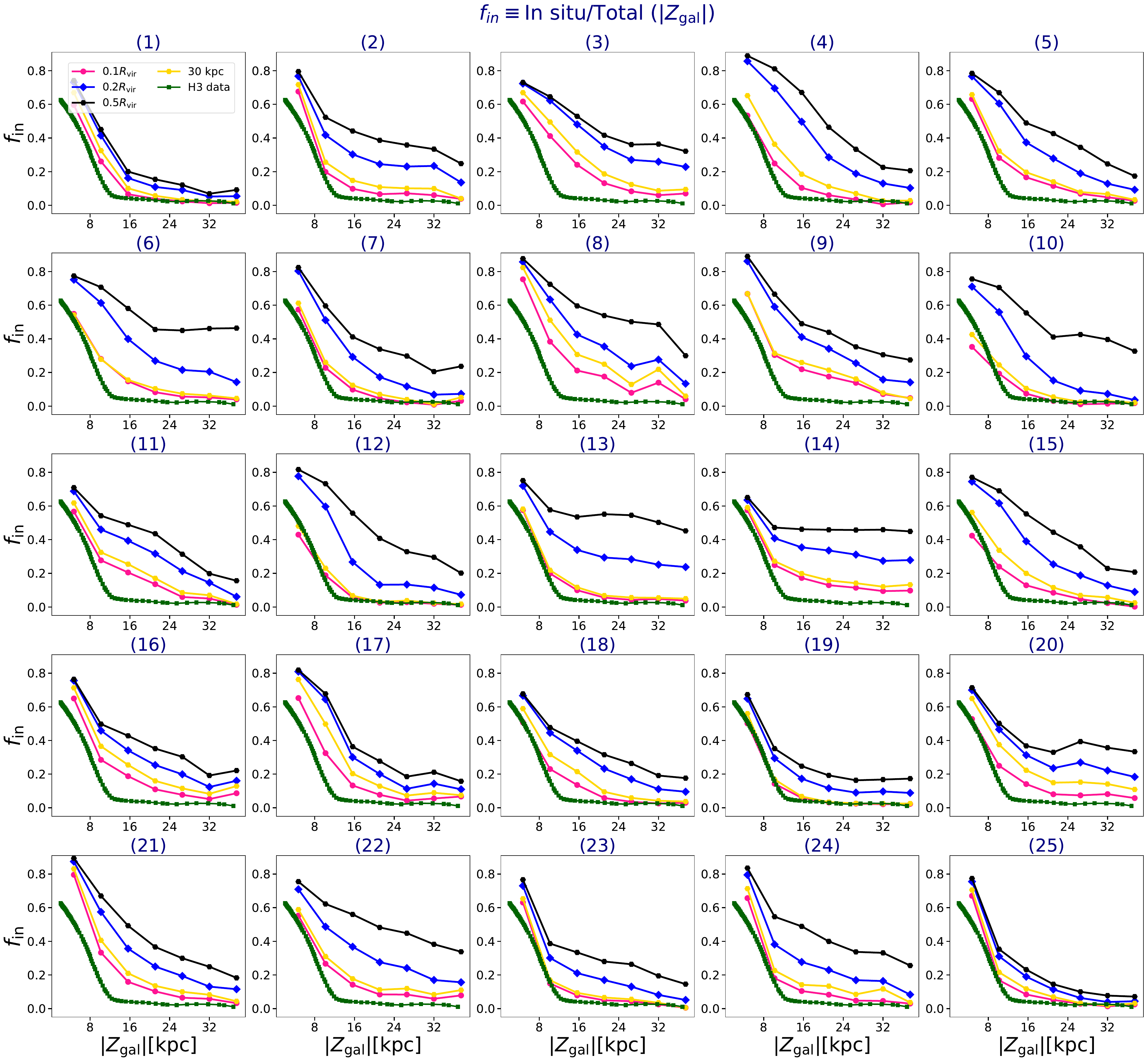}
\caption{The |Z| profile depicts the fraction of in~situ stars for different cutoffs, including (0.1, 0.2, 0.5) R$_{\mathrm{vir}}$ and 30 kpc, for a sample of 25 MW-like galaxies in the TNG50 simulation. In each panel, we have averaged over four different orientations for the Sun, located 8 kpc from the centre in the disk plane, spanning $\pm{\hat{i}}$ and $\pm{\hat{j}}$. This averaging process helps reduce noise from various orientations. Overlaid on each panel, the solid green line represents the results from the H3 survey. Notably, there is a fair general agreement between the TNG50 results and the H3 survey, particularly for lower radial cuts. However, the level of agreement diminishes with increasing the threshold radius.}
\label{Panel-insitu}
\end{figure*}

\subsection{The in~situ stars from TNG50 vs H3}
\label{Comparison-TNG-H3}
Having outlined our general methodology for inferring the in~situ stars from the TNG50 simulation, we proceed to compare our theoretical predictions with the actual observational results from the H3 survey.
\begin{figure*}
\center
\includegraphics[width=1.0\textwidth]{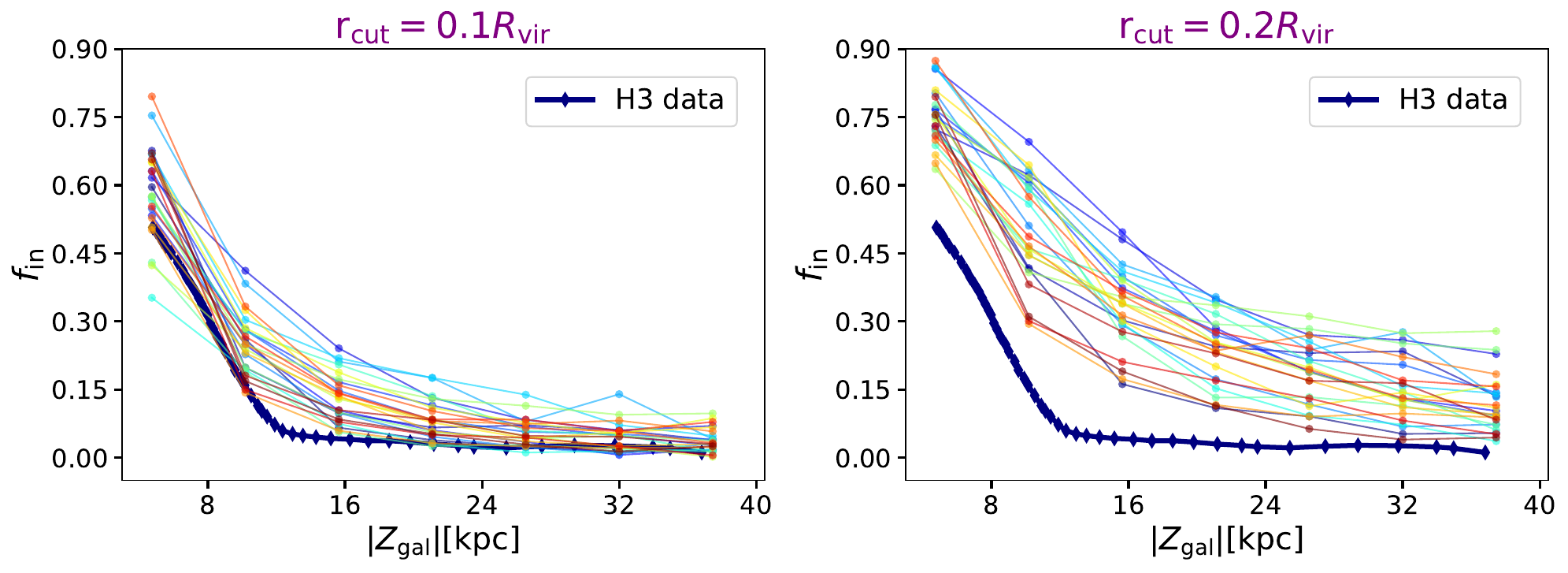}
\caption{The |Z$_{\mathrm{gal}}$| profile illustrates the fraction of in~situ stars to the total number of stars derived from the TNG50 simulation using two different radial cutoffs: $r{\mathrm{cut}} = 0.1 R_{\mathrm{vir}}$ (left) and $r_{\mathrm{cut}} = 0.2 R_{\mathrm{vir}}$ (right). Each panel encompasses the entirety of TNG50 results, with the observational results from the H3 survey overlaid on each diagram.}
\label{Collection-insitu01-02}
\end{figure*}
Figure \ref{Panel-insitu} illustrates the |Z| profile of the fraction of in~situ stars in various halos within our sample. Hereafter Z denotes the vertical height away from the disk. To establish the Z direction, we perform a coordinate transformation from the simulation frame to a coordinate system wherein the Z direction aligns with the total angular momentum of the stellar disk. Once we define the Z-direction, we maintain flexibility in selecting the new X-Y directions within the disk plane. Since the H3 survey provides results in the Heliocentric coordinate system, with the Sun located approximately 8 kpc away from the centre along the X direction, and given the absence of a real Sun in the TNG50 simulation, we repeat our analysis four times, placing the Sun in four orientations along the ($\pm{\hat{i}}$ and $\pm{\hat{j}}$) directions. Subsequently, we average the final results obtained from the different orientations.

In each panel, solid lines of different colours [pink, blue, black] represent results using $r_{\mathrm{cut}} = [0.1, 0.2, 0.5] R_{\mathrm{vir}}$, respectively, while the solid yellow line depicts results using $r_{\mathrm{cut}} = 30$ kpc. Additionally, the solid green line overlaid in each panel represents results from the H3 survey.

From the diagram, several observations can be made:

$\bullet$ Increasing the radial cutoff does not drive the ratio of in~situ stars to zero, indicating that in~situ stars are present throughout most galaxies.

$\bullet$ The |Z| profile of the ratio of in~situ stars varies among different galaxies. While some galaxies exhibit a significant decrease in $f_{\mathrm{in}}$ towards larger |Z|s, others display a smoother profile.

$\bullet$ Different galaxies exhibit varying sensitivity to different radial cutoffs. In some cases, increasing the radial cutoff notably enhances $f_{\mathrm{in}}$, whereas in others, it demonstrates less sensitivity.

$\bullet$ A constant cutoff of 30 kpc yields results similar to those obtained with a low radial cut of 0.1 $R_{\mathrm{vir}}$. This observation is intriguing as the curvature of the profile remains largely consistent across most cases, despite the former choice being a constant independent of individual galaxy specifics or the exact birth time of stars.

$\bullet$ In summary, a few galaxies [1, 11, 16, 18, 19, 23, 25] reasonably explain the trend in H3 results. Among these, only galaxies [1, 19, 25] match the observational results both in terms of the shape as well as the values of the in~situ fraction regardless of the particular spatial cut for inferring the in~situ fraction. 

Figure \ref{Collection-insitu01-02} displays the Z profile of the fraction of in~situ stars for all galaxies using two different radial cutoffs. The left panel corresponds to $r_{\mathrm{cut}} = 0.1 R_{\mathrm{vir}}$, while the right panel depicts $r_{\mathrm{cut}} = 0.2 R_{\mathrm{vir}}$. It is evident from the diagram that increasing $r_{\mathrm{cut}}$ leads to a greater deviation of the inferred in~situ fraction from the observational results obtained from the H3 survey. Conversely, at $r_{\mathrm{cut}} = 0.1 R_{\mathrm{vir}}$, most galaxies align well with the observational results. The lower cutoff appears more consistent with the constant cutoff used in previous literature \citep[see for example][]{2017ApJ...845..101B}.

Several factors may contribute to the discrepancy in the in-situ stellar fraction between H3 observations and TNG50 simulations. One possibility is numerical heating, which can artificially increase the scale height, leading to systematic deviations between theoretical predictions and observations. Alternatively, the observed differences may stem from variations in the merger histories of individual halos. Acknowledging these potential systematic effects as limitations of our comparison, we proceed in the following sections to investigate the relationship between the deviation in the in-situ stellar fraction from the H3 results and the merger histories of the galaxies in our sample.

\begin{figure*}
\center
\includegraphics[width=1.0\textwidth]{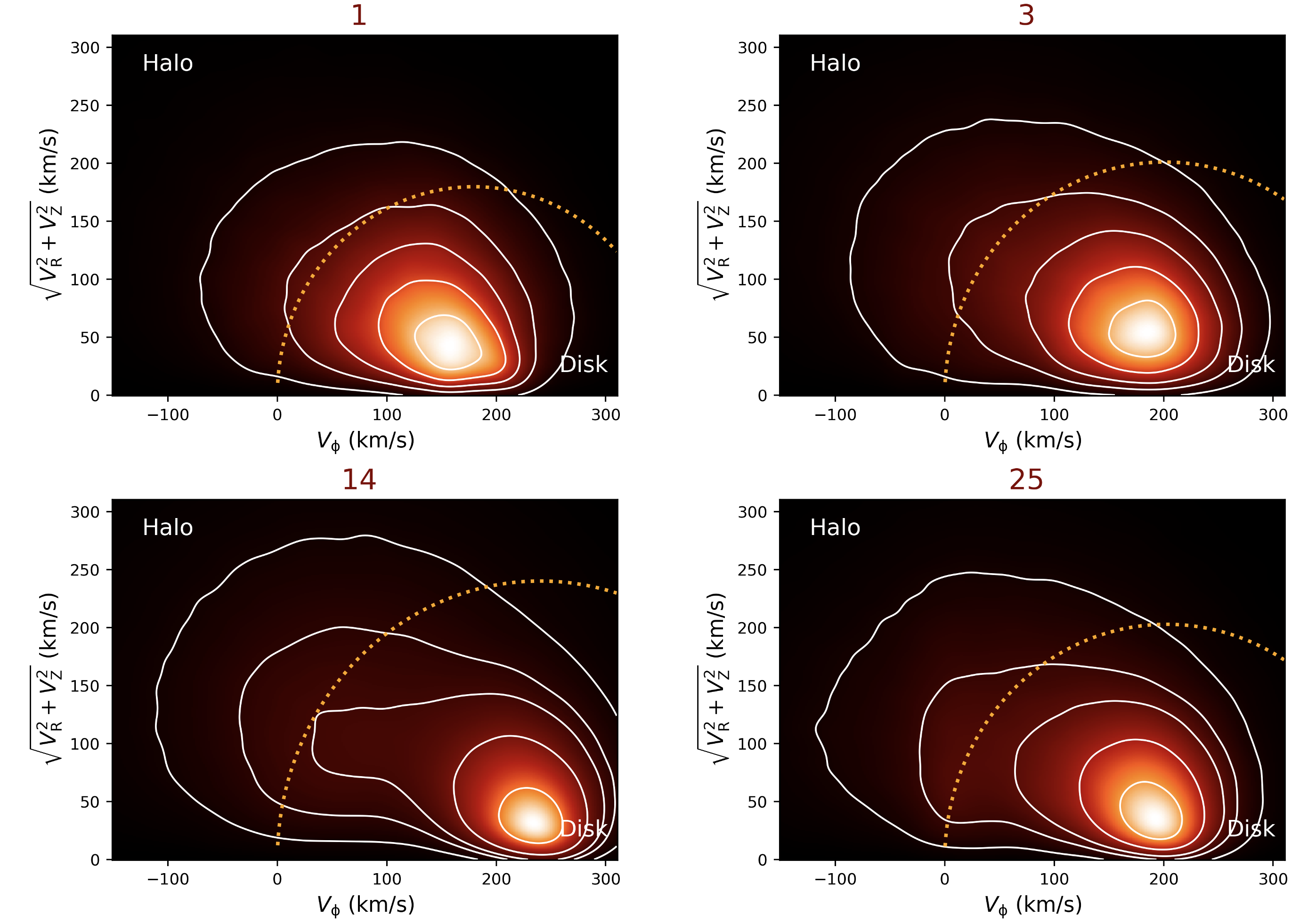}
\caption{The Toomre diagram depicting the distribution of stars in four galaxies from our MW-like galaxy sample. 
The white contours in each panel (from the inner part to the outer part) refer to the top 10\%, 30\%, 50\%, 70\%, and 90\% of stars, respectively. It is inferred that in each case there is a tail of stars on halo orbits, while the majority of stars are on the disk orbits. The dotted orange line in each panel shows the boundary between the halo and disk stars.
}
\label{Toomere-Diagram}
\end{figure*}

\subsection{On the Origin of in~situ stars}
\label{Insitu-origin}
There have been various hypotheses proposed regarding the origin of in~situ stars in galaxies. \cite{2010MNRAS.403.1283G, 2018MNRAS.474..443G} suggested that the in~situ component might originate from the collapse of cold gas, which is brought in from numerous gas-rich mergers. Conversely, they proposed that accreted stars could form from stellar sub-halos accreted onto the host galaxy. However, recent advancements, particularly through the analysis of Gaia DR1 and DR2, have significantly expanded our understanding beyond the solar neighbourhood, providing insights into the galactic halo as well. Gaia data suggests that the in~situ component of the galactic halo might be associated with a population of heated disk stars \citep[see for example][and references therein]{2017ApJ...845..101B, 2018ApJ...863..113H, 2019A&A...632A...4D, 2020MNRAS.494.3880B, 2019MNRAS.484.2166S}. These observational findings align well with N-body analyses \citep[see for instance][and references therein]{2010ApJ...721..738Z, 2010MNRAS.404.1711P, 2011MNRAS.416.2802F, 2011A&A...535A...5Q, 2012MNRAS.420.2245M, 2017A&A...604A.106J}.

The Toomre diagram is a powerful tool for distinguishing different stellar components of the Milky Way, particularly between the stellar disk and halo populations \citep[see for example][and references therein]{2017ApJ...845..101B, 2020A&A...636A.115D}. Disk stars typically exhibit prograde motion with azimuthal velocities close to the Local Standard of Rest (LSR) (defined below), and their density gradually decreases as velocities deviate from the LSR, eventually reaching retrograde speeds. Since the relative number densities of disk and halo stars are shaped by past galaxy mergers, the Toomre diagram serves as an effective diagnostic for identifying merger remnants. When combined with metallicity and chemical abundance data, it provides a deeper link between stellar kinematics and population properties, offering insights into the formation history and evolutionary process of the Milky Way.
Motivated by these insights, we use the Toomre diagram to investigate the redshift evolution of stars in various orbits, encompassing both members of the stellar halo and the stellar disk and investigate the role of galaxy mergers in altering the relative number of disk versus halo stellar components. As a part of our analysis, we construct the Toomre diagram for a subset of galaxies in our MW-like galaxy sample at redshift zero. Subsequently, we leverage this diagram across the entire redshift evolution to classify stars into halo-like and disk-like components. Furthermore, we explore the role of galaxy mergers in influencing the evolution of the halo-like and disk-like components. 

Figure \ref{Toomere-Diagram} illustrates the Toomre diagram at redshift zero for a subset of 4 galaxies in our sample, with all stars being included in the diagram. Each panel displays an overlaid dashed line representing the boundary between the stars in the disk-like and those on the halo-like orbits, defined as:

\begin{align}
\label{disk-halo}
|V - V_{\mathrm{LSR}}| \geq & V_{\mathrm{LSR}} ~,~~~~~~~~~ \mathrm{Halo} \nonumber\\
 |V - V_{\mathrm{LSR}}| < & V_{\mathrm{LSR}} ~,~~~~~~~~~
 \mathrm{Disk}
\end{align}
where we define $V_{\mathrm{LSR}}$ as the local standard of rest (LSR) speed, which is equivalent to the circular speed, $V_{\mathrm{cir}}(r)$, at the location of the Sun. Here, $V_{\mathrm{cir}}(r) \equiv \sqrt{G M_{\mathrm{tot}}(r)/r}$, where $M_{\mathrm{tot}}(r)$ refers to the total mass interior to distance $r$ including DM, star and gas particles. We consider the sun to be located at 8 kpc away from the galactic centre at redshift zero. The x-axis of the Toomre diagram depicts the azimuthal component of individual stars. It is inferred in two steps. First, we compute the total angular momentum of all of the stars. We subsequently infer the alignment of the angular momentum of each stellar particle with respect to this vector, as well as the component of their location along this direction. $V_{\phi,i}$  is then given as: 
\begin{equation}
V_{\phi,i} = j_{z,i}/\left(m_iR_{\mathrm{perp},i}\right)~~~, ~~~ R_{\mathrm{perp},i} \equiv \sqrt{r_i^2 - z_i^2}
\end{equation}
For convenience, we have removed the sub-index $i$ in Figure \ref{Toomere-Diagram}.
while the y-axis showcases the velocity orthogonal to the stellar disk angular momentum.

overlaid in Figure \ref{Toomere-Diagram}, we have drawn contours representing the top 10\%, 30\%, 50\%, 70\%, and 90\% of stars, ordered from the innermost to the outermost regions. It is evident that, across all galaxies, the top 90\% contour reveals a tail of stars transitioning into halo orbits. However, the prominence of this tail at lower percentile levels varies depending on the individual galaxy. For instance, in galaxy 1, the Toomre diagram is highly compact, with nearly all stars—up to the 90\% contour—remaining in the disk. In contrast, galaxy 14 exhibits the most extended stellar tail in halo orbits, including a population with retrograde motion. Galaxies 3 and 25 present intermediate cases, where a significant number of stars are distributed in both the disk and halo. 

The selection of these galaxies is based on their in~situ fraction profiles from Figure \ref{Panel-insitu}. Galaxies 1 and 25 exhibit profiles closely resembling the H3 observations, while galaxies 3 and 14 show progressively greater deviations from the observations as the spatial cut increases. Galaxy 1 exhibits a highly compact Toomre diagram, consistent with its strong similarity to the H3 survey and the presence of very few stars on halo orbits. In contrast, galaxy 14 displays the most extended Toomre diagram, with a prominent tail of stars in halo orbits. This trend aligns with the increased deviation between the in~situ stellar fraction in TNG50 and the H3 survey at larger spatial cuts. Interestingly, the Toomre diagrams of galaxies 3 and 25 appear very similar, despite notable differences in their in~situ stellar fraction trends. In the case of galaxy 3, deviations emerge at larger spatial cuts, whereas in galaxy 25, the agreement between theory and observation remains robust regardless of the spatial cut used to define in-situ stars. This distinction is reasonable, as the Toomre diagram represents stellar distribution in velocity space, while the in~situ stellar fraction is defined based on spatial coordinates. Given the possible limitation of connecting these two space, moving forward we will use alternative metrics to quantify the observed deviations between the MW-like galaxies from the TNG50 simulations and the H3 observation. 

\begin{figure}
\center
\includegraphics[width=0.45\textwidth]{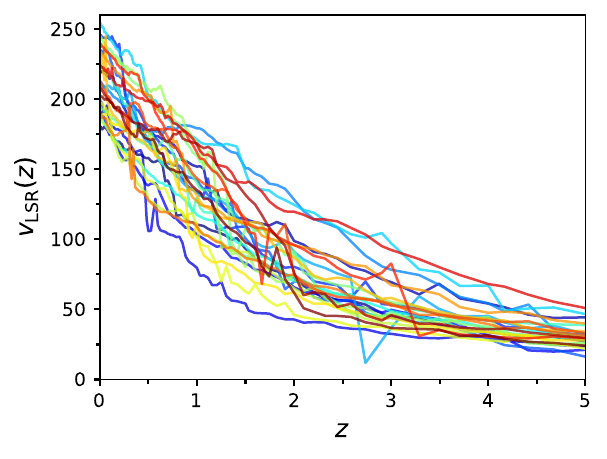}
\caption{The redshift evolution of the local standard of rest velocity for 25 MW-like galaxies in our sample. The velocity is in the unit of km/s.}
 \label{Toomere-z}
\end{figure}
As $V_{\mathrm{LSR}}$ speed is essential in the distinction between the halo-like and disk-like stars, in Figure \ref{Toomere-z} we present the redshift evolution of this quantity, defined as the circular speed $V_{\mathrm{cir}}$ evaluated at $r = A_{0} R_{\mathrm{h}}(z)$. Here, $R_{\mathrm{h}}(z)$ represents the redshift-dependent half-light radius, and $A_{0}$ is a constant determined such that $r = 8$ kpc at redshift zero. Thus, we have $A_{0} = 8/R_{\mathrm{h}}(z=0)$. The choice of 8 kpc ensures that the $V_{\mathrm{LSR}}$ is computed at the Sun's location at present. 

The key takeaway from Figure \ref{Toomere-z} is the universality of the $V_{\mathrm{LSR}}$ as a function of redshift for different galaxies in our galaxy sample.
\begin{figure*}
\center
\includegraphics[width=1.0\textwidth]{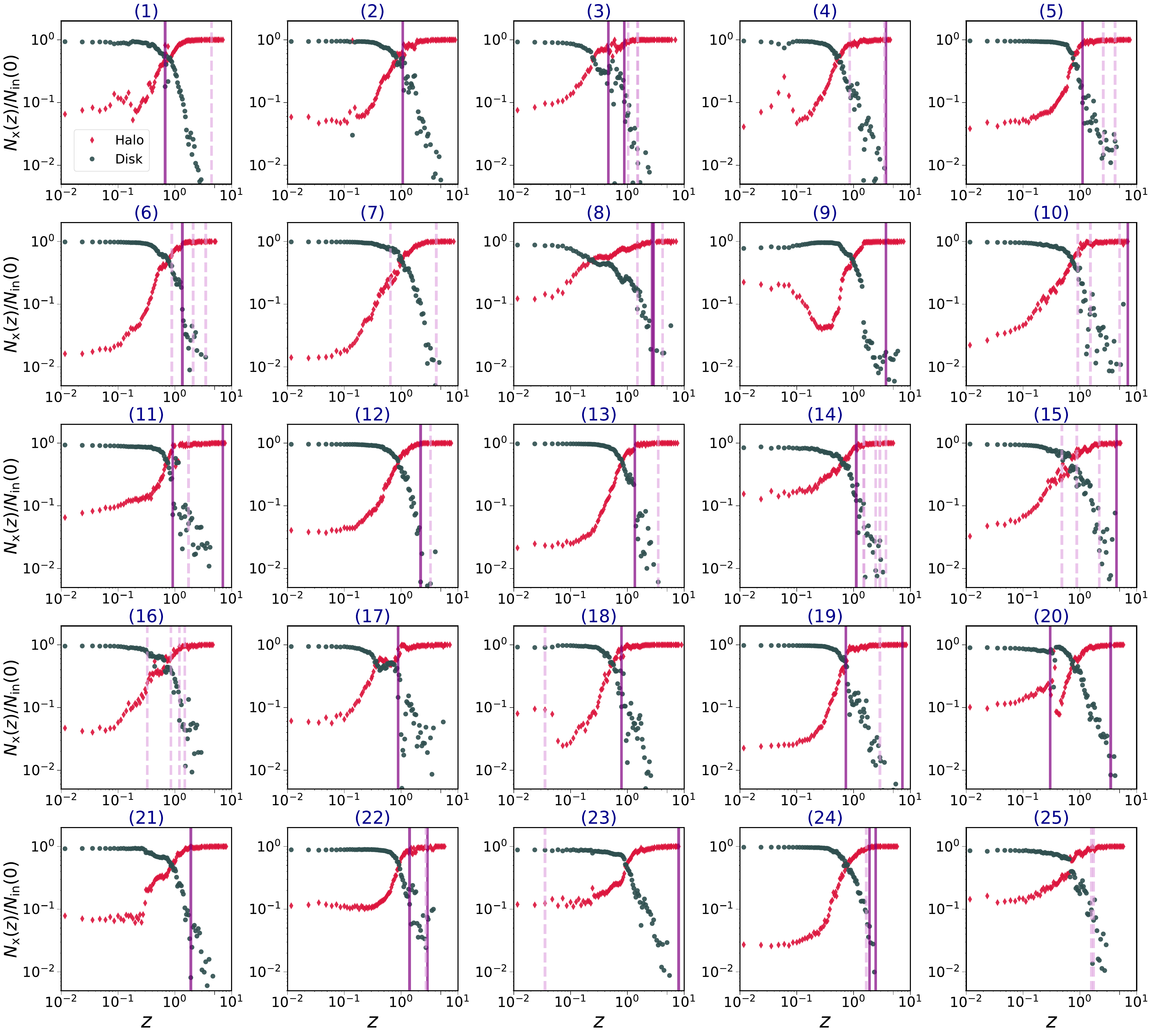}
\caption{The redshift evolution of the fraction of the number of disk (represented by grey-circle) and halo (depicted by red-diamond) in~situ stars between $r = 17-23$ kpc to the total number of in~situ stars at redshift zero with a cutoff of 0.2 $R_{\mathrm{vir}}$. Additionally, minor and major mergers are indicated by dashed-plum and solid-purple lines, respectively. The thresholds for minor and major mergers are set as 0.05-0.2 and above 0.2, respectively. It is seen that both the major and minor mergers change the slope of the evolution of the fraction of in~situ stars. }
\label{Insitu-Redshift}
\end{figure*}

Figure \ref{Insitu-Redshift} illustrates the redshift evolution of the ratio of halo (denoted by diamond-red markers) and disk (represented by grey-circle markers) in~situ stars as identified at redshift zero defined using spatial cut at 0.2 $R_{\mathrm{vir}}$. It is important to note that the choice of 0.2$R_{\mathrm{vir}}$ is motivated by the need to include a larger number of stars while effectively capturing the impact of galaxy mergers on the orbits of both disk  and halo stars.
While the in~situ stars are defined using a spatial cut at 0.2 $R_{\mathrm{vir}}$, we only show the evolutionary trajectory of their number at $r = 17-23$ kpc at zero redshifts, to the total number of in~situ stars at redshift zero within the same radial range. We have chosen this range, $r = 17-23$ kpc, as according to Figure \ref{Panel-insitu}, the in~situ stellar fraction diminishes significantly in this interval. Furthermore, to simplify delivering the picture, we only present the results for a case with the Sun located at 8 kpc in the +X direction.

\begin{figure*}
\center
\includegraphics[width=0.49\textwidth]{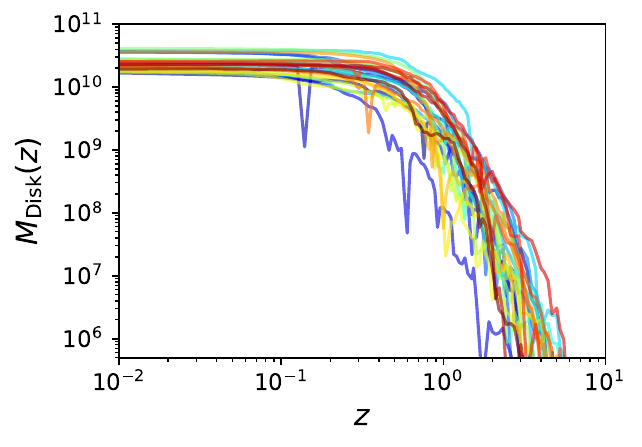}
\includegraphics[width=0.49\textwidth]{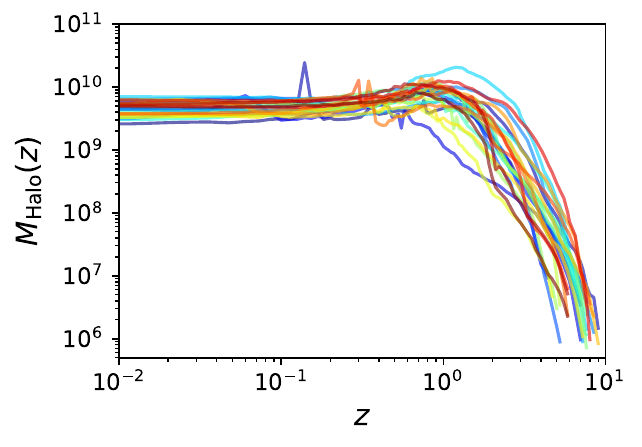}
\caption{The redshift evolution of the total mass of the disk in~situ stars (left panel) and halo in~situ stars (right panel), defined with a spatial cutoff of 0.2 $R_{\mathrm{vir}}$.}
 \label{Mass-Insitu}
\end{figure*}
For each galaxy, we begin with the selected in~situ stars at zero redshift, identified based on the criteria outlined in Section \ref{Comparison-TNG-H3} and trace them back in time. At each redshift, we compute the $V_{\mathrm{LSR}}$ and categorise the stars into halo (diamond-red) and disk (grey-circle) components. Additionally, major mergers are depicted by solid-purple lines, while minor mergers are represented by dashed-plum lines.

The two ratios are complementary as expected. It is evident that stars on disk orbits begin with a negligible fraction at high redshifts and gradually increase in their percentage towards lower redshifts. Ultimately, at  redshift zero, the in~situ stellar fraction is predominantly dominated by stars in disk-like orbits, which aligns with the predominant stellar disk characteristic of MW-like galaxies. Galaxy mergers play a role in altering the growth/decline slope for both halo and disk stars.

Figure \ref{Mass-Insitu} illustrates the redshift evolution of the in~situ stellar mass in disk orbits (left panel) and halo orbits (right panel). The original in~situ stars are traced back in time, and in each snapshot, the stellar mass in the disk and halo orbits is inferred. It is observed that throughout the galaxy's evolution, the total stellar mass in both disk and halo orbits increases, with the final mass in disk stars being dominant over their values in halo orbits.

\subsection{Redshift Evolution of R$_{\mathrm{vir}}$ and SFR}
\label{SFR-Rvir}

\begin{figure*}
\center
\includegraphics[width=1.0\textwidth]{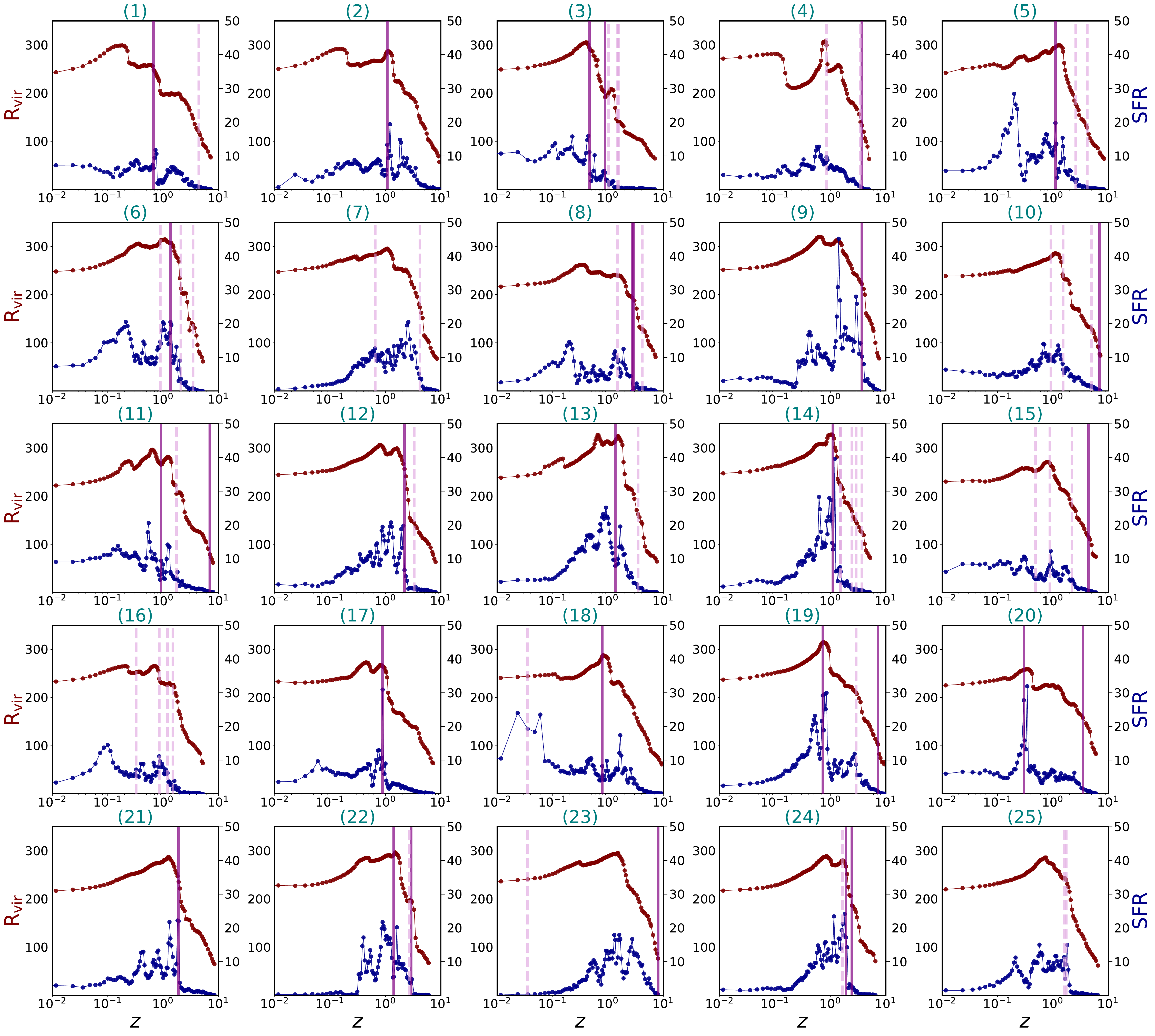}
\caption{Redshift evolution of the comoving virial radius (R${\mathrm{vir}}$) (left) and the star formation rate (SFR) (right) for our sample of MW-like galaxies. Vertical solid lines (purple) indicate major mergers, while dashed lines (plum) represent minor mergers. It is observed that both R${\mathrm{vir}}$ and SFR are enhanced during merger events.}
\label{Panel-Rvir-SFR}
\end{figure*}

With the merger history of the first progenitor elucidated, we now delve into the redshift evolution of two key parameters in our analysis: the comoving virial radius $R_{\mathrm{vir}}$ and the star formation rate (SFR). The comoving virial radius is pertinent as it influences the radial cutoff used to infer the in~situ stars at each redshift. On the other hand, the SFR is linked to the in~situ stellar budget at any given time. Both of these quantities are impacted by merger events.

Figure \ref{Panel-Rvir-SFR} presents the redshift evolution of $R_{\mathrm{vir}}$ and the SFR for our entire MW-like sample. In each diagram, the time of major mergers is indicated by a solid purple line, while minor mergers are represented by dashed plum lines.

Throughout the galaxy's evolutionary trajectory, instances arise where either or both $R_{\mathrm{vir}}$ and SFR change their amplitude and slope. While there are various reasons for these slope variations, e.g. owing to galaxy-galaxy mergers, due to pseudo-evolution owing to cosmic evolution, below we mainly focus on their impact on the evolution of in~situ star fraction. More specifically, we embark on an analysis to explore the correlation between the evolution of these quantities and the deviation between the in~situ fraction inferred from the TNG50 simulations and the H3 survey. 

\subsubsection{Quantitative comparison between TNG50 and H3 results} \label{Quantitative-Comp}
Here we construct a more quantitative metric to facilitate the comparison between the scale-height distribution of the in~situ stars from the H3 survey and the TNG50. It is noted from each panel in Figure \ref{Panel-insitu} that the in~situ fraction from H3 survey has two important features. Its sharp slope at smaller $\vert Z \vert$, i.e.  $\vert Z \vert: (5-7.5)$kpc, followed by its amplitude decline at larger distances, i.e. $\vert Z \vert: (17-20)$kpc. To make an in-depth comparison between the theory and observation we include both of the amplitude and the slope (i.e. the derivative of the in~situ star fraction)
inside our metric. We have made a dedicated test analysis of either dropping the amplitude difference or the slope difference from the metric and realised that in both cases the strength of the subsequent correlation functions (as will be extensively discussed in Section \ref{Difference-TNG-H3}) with both of the internal and external drivers diminishes. Having this pointed out, we suggest the following quantitative metric for comparing the H3 survey with TNG50:

\begin{align}
\label{Effective-Rank}
\Delta f^{j}_{\mathrm{eff}} \equiv & ~ 
\bigg{\vert} \frac{ \sum _{i = 1}^{\mathrm{N_1}} \bigg{(} f^{j}_{\mathrm{in}}(i) - f_{\mathrm{H3}}(i) \bigg{)} }{2\mathrm{N_1}} \bigg{\vert}  + \bigg{\vert} \frac{
\sum _{i = 1}^{\mathrm{N_2}} \bigg{(} f'^{j}_{\mathrm{in}}(i) -   f'_{\mathrm{H3}}(i) 
\bigg{)}}{\mathrm{N_2}} \bigg{\vert} \nonumber\\
\equiv & ~ \Delta f^{j}_{\mathrm{Amp}}  + \Delta f^{j}_{\mathrm{Slope}} ~~~,~~~ j = (0.1, 0.2, 0.5).  
\end{align}
 
here $f^{j}_{\mathrm{in}}(i)$ refers to the amplitude of the fraction of the in~situ stars from the TNG50. The upper index describes the spatial cut in inferring the in~situ stars from simulations, while the sub-index $i$ presents the scale height that we have evaluated the in~situ star fraction as outlined below. $f_{\mathrm{H3}}(i)$ describes the fraction of the in~situ stars from the H3 survey located in $i$th bin. 
We split the scale height between $\vert Z \vert: (17-20)$kpc to $N_1 = 23$ linear bins and evaluate the difference between the TNG50 and the H3 in each of these points. We then compute the mean value of the amplitude difference including all of these $N_1$ points.

In the second term we infer the difference between the derivative of the in~situ fraction with respect to the scale height. Here we cover the scale height in the range 5-7.5 kpc and split this range to $N_2$= 10 points.
$f'^{j}_{\mathrm{in}}(i)$ refers to the derivative of the fraction of in~situ star defined with the spatial cut at $j$ with respect to the scale height, while $f'_{\mathrm{H3}}(i)$ describes the derivative of the in~situ fraction from H3 survey evaluated at the bin $i$th.

Since the theoretically inferred in~situ fraction depends on the actual cutoff, we may want to try different deviations at the level of 0.1$R_{\rm{vir}}$, 0.2$R_{\rm{vir}}$, and 0.5$R_{\rm{vir}}$. 
Using $\Delta f^{j}_{\mathrm{eff}}$, we can associate one number to every galaxy and use it as a proxy to judge how close the simulated galaxy is to the observations. Another advantage of this metric is that we can make an automated way to accurately infer the proximity between the theory and H3 observation. 

Below, we extensively make use of  $\Delta f^j_{\mathrm{eff}}$ when we connect the above deviations to the merger history and the actual properties of individual halos. Our goal is to find some possible correlations between how small/large is $\Delta f^j_{\mathrm{eff}}$ and the closeness of different MW-like galaxies chosen from the TNG and the actual MW from the H3. 

\section{Exploring different drivers for the in~situ stars}
\label{Difference-TNG-H3} 
Having quantified the effective deviation in the in~situ stellar fraction between the TNG50 and H3 survey datasets, our objective is to establish a correlation between this metric and the key parameter characterising each galaxy, as well as its assembly history. This analysis involves two main categories of parameters: internal drivers and external factors. Below, we investigate the influence of these drivers and their association with the $\Delta f^j_{\mathrm{eff}}, j = (0.1, 0.2, 0.5)$ quantity derived from Equation \ref{Effective-Rank}.

\begin{figure*}
\center
\includegraphics[width=0.88\textwidth]{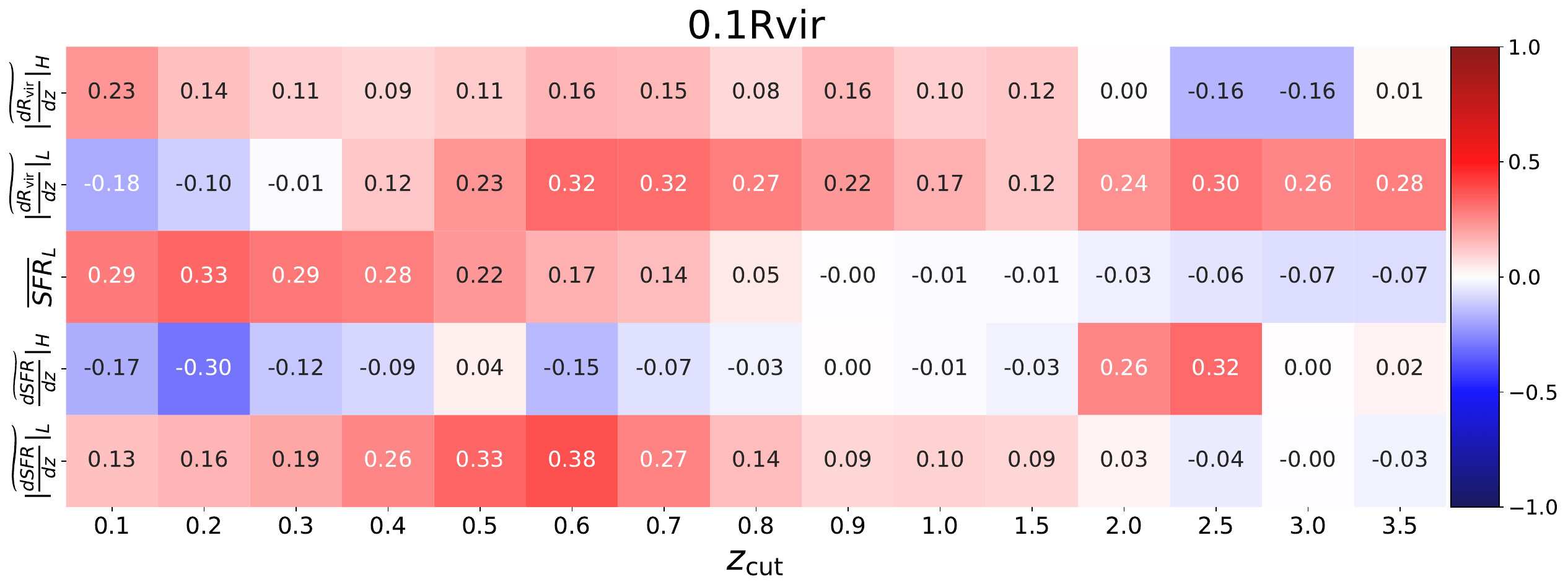}
\includegraphics[width=0.88\textwidth]{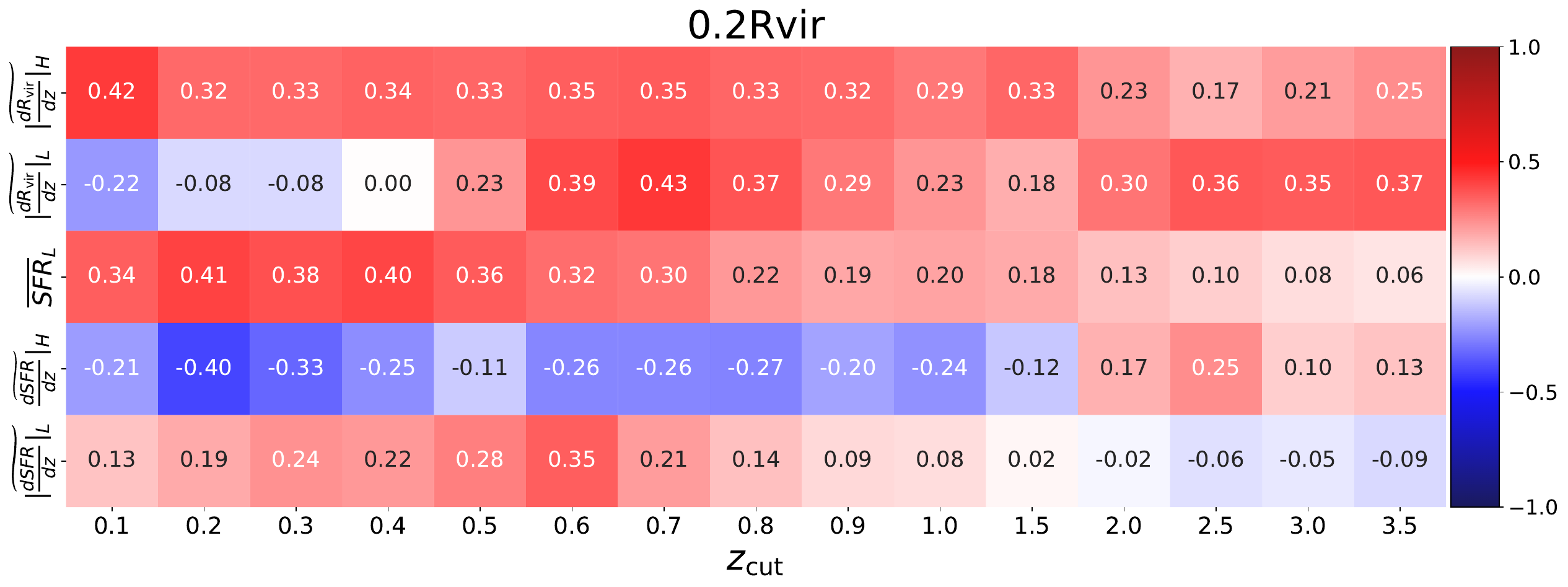}
\includegraphics[width=0.88\textwidth]{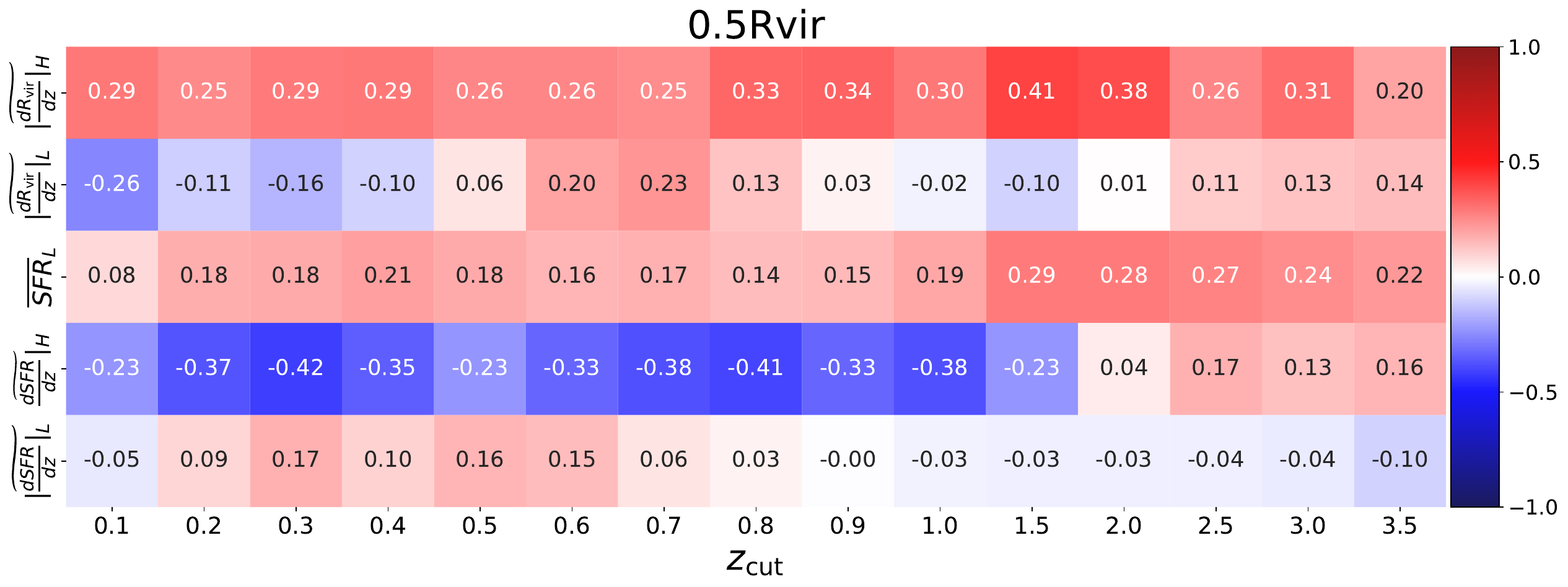}
\caption{The correlation coefficient of 
$\widetilde{|\frac{dR_{\mathrm{vir}}}{dz}|}_H$, $\widetilde{|\frac{dR_{\mathrm{vir}}}{dz}|}_L$, $\overline{\mathrm{SFR}}_L$, $\widetilde{\frac{d \mathrm{SFR}}{dz}}|_H$ and  $\widetilde{|\frac{d \mathrm{SFR}}{dz}|}_L$ with $\Delta f^j_{\mathrm{eff}}$. Different matrices from top to bottom correspond to various spatial cuts of  $j = (0.1, 0.2, 0.5)$, respectively.} 
\label{correlation-matrix-internal}
\end{figure*}

\subsection{Internal drivers for the in~situ fraction}
\label{Internal-drivers}
As discussed in Section \ref{SFR-Rvir}, $R_{\mathrm{vir}}$ and the SFR are fundamental internal parameters with direct implications for galaxy evolution. Building upon this insight, we investigate the correlation matrix between these parameters and the deviation between the in~situ fraction from the TNG50 simulation and H3 survey as given in  Equation \ref{Effective-Rank}. Given the evolutionary nature of these internal drivers over a galaxy's lifespan, we segment the data into redshift bins to analyse their significance at different epochs. Specifically, we compute the mean, median, and slope of these parameters using two distinct approaches: either from an initial redshift to $z_{\mathrm{cut}}$ (referred to as $H$), or from $z_{\mathrm{cut}}$ to redshift zero (named as $L$). Subsequently, we assess the correlation between these metrics and $\Delta f^j_{\mathrm{eff}}, j = (0.1, 0.2, 0.5)$ derived from Equation \ref{Effective-Rank}. 
  
We conduct an extensive exploratory analysis to assess the significance of different combinations and operations on  $R_{\mathrm{vir}}$ and $\mathrm{SFR}$ in our study, constructing correlation matrices using both the mean and median of these parameters, as well as their derivatives. We also incorporate both the original and the absolute-valued derivatives. \\
Within each correlation matrix, the relative importance of different metrics fluctuates across  $z_{\mathrm{cut}}$  values, highlighting the role of both amplitude and slope in capturing deviations between theory and observation. Notably, the mean  $\mathrm{SFR}$  exhibits a strong correlation, while the slope of  $R_{\mathrm{vir}}$  proves to be more significant than its value. Furthermore, the median slope of both  $\mathrm{SFR}$  and  $R_{\mathrm{vir}}$  emerges as more influential than their mean counterparts, suggesting that gradual variations in these quantities are more impactful than outliers.  \\
Interestingly, for  $\frac{dR_{\mathrm{vir}}}{dz}$, the absolute-valued median exhibits stronger correlations, whereas for  $\mathrm{SFR}$  and  $\frac{d \mathrm{SFR}}{dz} \Big|_H$ , the mean and median absolute values show weaker correlations. This finding led us to adopt a flexible approach, selecting either the mean or median, as well as either the parameter value or its slope in our analysis.  \\ 
Having this pointed out, we keep the following discussions concise and present only the most significant parameter combinations in our correlation matrix.
This includes:  $\widetilde{\frac{dR_{\mathrm{vir}}}{dz}}|_H$, $\widetilde{\frac{dR_{\mathrm{vir}}}{dz}}|_L$, $\overline{SFR}_L$, $\widetilde{\frac{d SFR}{dz}}|_H$, and  $\widetilde{\frac{d SFR}{dz}}|_L$, The median and mean of the X quantity are denoted by $\tilde{\mathrm{X}}$ and $\overline{\mathrm{X}}$, respectively.

Figure \ref{correlation-matrix-internal} illustrates the correlation coefficient matrix for both the $H$ and $L$ redshift segments across varying $z_{\mathrm{cut}}$ values. Each row represents a spatial cut, progressively increasing from $0.1 R_{\mathrm{vir}}$ to $0.5 R_{\mathrm{vir}}$. While there are discernible variations in correlation coefficients between the smallest ($0.1 R_{\mathrm{vir}}$) and largest ($0.5 R_{\mathrm{vir}}$) cuts, overall trends, including magnitude and sign, remain consistent. 

\begin{figure*}
\center
\includegraphics[width=0.96\textwidth]{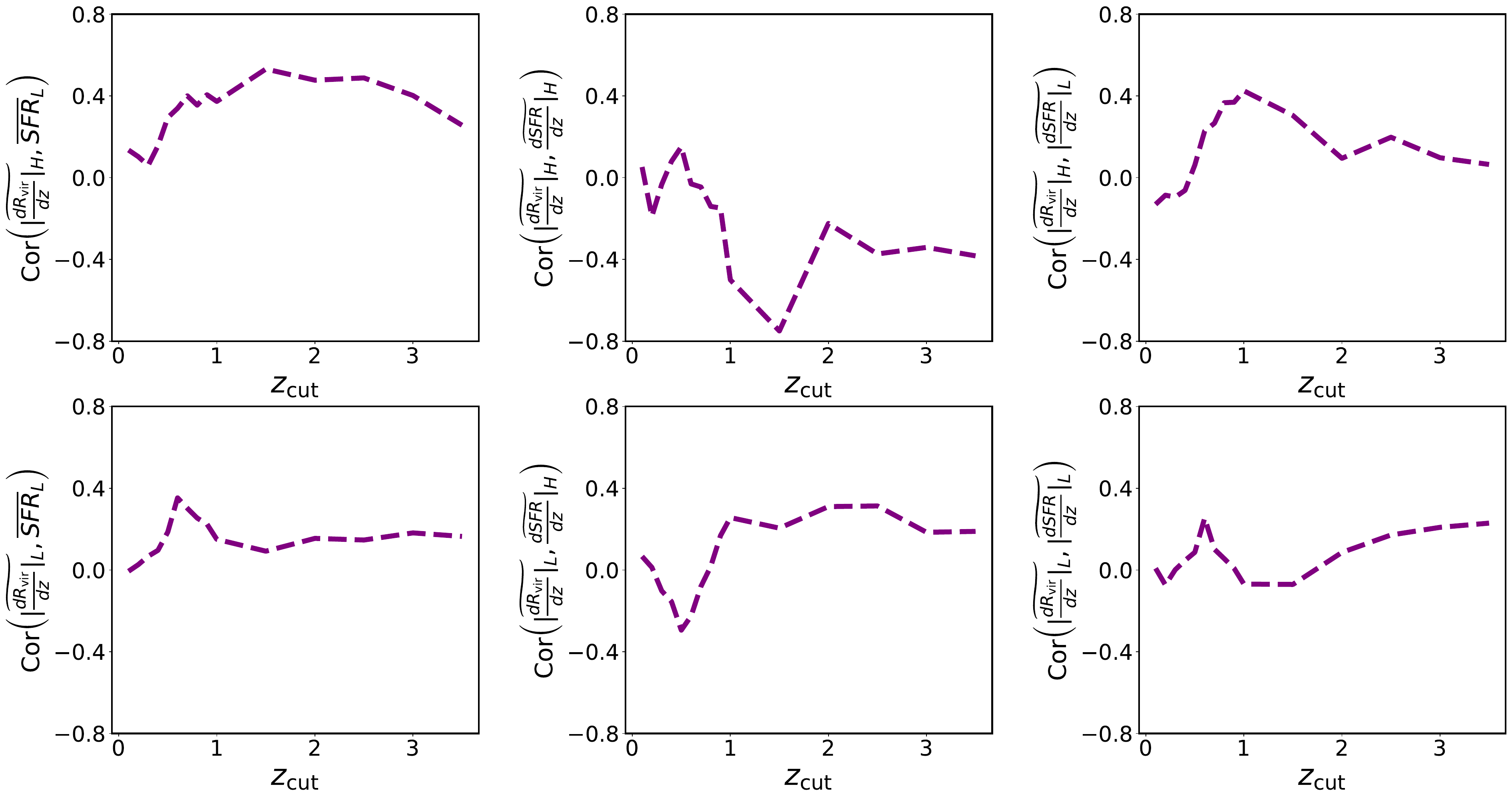}
\caption{The cross-correlation coefficient between different internal metrics. In top/bottom row, from left to right, we present the correlation of $\widetilde{\frac{dR_{\mathrm{vir}}}{dz}}|_H$/$\widetilde{\frac{dR_{\mathrm{vir}}}{dz}}|_L$ with $\overline{SFR}_L$, $\widetilde{\frac{d SFR}{dz}}|_H$ and  $\widetilde{\frac{d SFR}{dz}}|_L$, respectively. }
 \label{Cross-Correlation-internal}
\end{figure*}

It is evident that in most cases, $\widetilde{|\frac{dR_{\mathrm{vir}}}{dz}|}_H$ is positively correlated with $\Delta f^j_{\mathrm{eff}}, j = (0.1, 0.2, 0.5)$. To elucidate the underlying reason for this positive correlation, we delve deeper into Figure \ref{Panel-Rvir-SFR}, where it becomes apparent that $R_{\mathrm{vir}}$ undergoes two distinct evolutionary phases: a growth phase followed by a contracting phase. The transition point from the former to the latter phase varies across different galaxies. An earlier rapid growth expands the boundaries of $R_{\mathrm{vir}}$ to larger distances, consequently increasing the fraction of in~situ stars and thereby elevating the value of $\Delta f^j_{\mathrm{eff}}, j = (0.1, 0.2, 0.5)$. Moreover, Figure \ref{correlation-matrix-internal} suggests that for a very early redshift cut, i.e., the higher values of $z_{\mathrm{cut}}$, larger spatial cuts are more correlated than the lower values. We posit that this phenomenon is associated with the inclusion of the transitional phase from an increasing to diminishing evolutionary phase of $R_{\mathrm{vir}}$.

At lower values of $z_{\mathrm{cut}}$, $\widetilde{|\frac{dR_{\mathrm{vir}}}{dz}|}_L$ is anti-correlated with $\Delta f^j_{\mathrm{eff}}, j = (0.1, 0.2, 0.5)$, indicating that as $R_{\mathrm{vir}}$ diminishes, the spatial boundaries contract, resulting in a more concentrated distribution of stars and thus lower values of $\Delta f^j_{\mathrm{eff}}, j = (0.1, 0.2, 0.5)$. Conversely, at higher values of $z_{\mathrm{cut}}$, this anti-correlation transitions to a positive correlation. We interpret this shift as a consequence of changes in the slope of $R_{\mathrm{vir}}$ corresponding to increasing $z_{\mathrm{cut}}$. 

Figure \ref{correlation-matrix-internal} illustrates a positive correlation between $\overline{\mathrm{SFR}}_L$ and $\Delta f^j_{\mathrm{eff}}, j = (0.1, 0.2, 0.5)$. This correlation signifies the potential for higher $\overline{\mathrm{SFR}}_L$ values to stimulate star formation across various regions of galaxies. Notably, the strength of this correlation fluctuates with $z_{\mathrm{cut}}$: stronger correlations are evident at smaller spatial cuts for lower $z_{\mathrm{cut}}$ values, while stronger correlations occur at larger spatial cuts for higher $z_{\mathrm{cut}}$ values. We propose that this behaviour is driven by the evolutionary dynamics of galaxies. Due to the higher surface density and the Schmidt-Kennicutt relation, stars are more likely to be concentrated near the centre at low redshift, resulting in stronger correlations for smaller spatial cuts. In contrast, at higher redshifts, star formation occurs at larger distances, thereby enhancing correlations at larger spatial cuts.

Comparing the correlation coefficients of $ \widetilde{\left| \frac{dR_{\mathrm{vir}}}{dz} \right|}_H $ with $ \widetilde{\frac{d \mathrm{SFR}}{dz}}|_H $, it is evident that they closely mirror each other in most cases. The key distinction lies in the former being the median of the absolute values, whereas the latter is simply the median. This difference leads to a reversal in the sign of the correlation coefficient: $ \widetilde{\left| \frac{dR_{\mathrm{vir}}}{dz} \right|}_H $ exhibits only positive correlations, while $ \widetilde{\frac{d \mathrm{SFR}}{dz}}|_H $ predominantly shows negative correlations. Despite this sign difference, their correlation magnitudes remain fairly similar.

In summary, the correlation coefficient structure exhibits remarkable consistency between $R_{\mathrm{vir}}$ and SFR. In most cases, the median of the absolute valued derivative shows a positive correlation with $\Delta f^j_{\mathrm{eff}}, j = (0.1, 0.2, 0.5)$, underscoring the role of enhancements in these values in amplifying the in~situ stellar fraction across galaxies.

Figure \ref{Cross-Correlation-internal} showcases the cross-correlation between various internal metrics and $z_{\mathrm{cut}}$. The top and bottom rows correspond to $\widetilde{\frac{dR_{\mathrm{vir}}}{dz}}|_H$ and $\widetilde{\frac{dR{\mathrm{vir}}}{dz}}|_L$, respectively, in conjunction with stellar-driven metrics. Within each row, from left to right, we demonstrate the correlation of $\widetilde{|\frac{dR{\mathrm{vir}}}{dz}|}_{i = (H,L)}$ with $\overline{\mathrm{SFR}}_L$, $\widetilde{\frac{d \mathrm{SFR}}{dz}}|_H$, and $\widetilde{|\frac{d \mathrm{SFR}}{dz}|}_L$, respectively. The presence of non-zero cross-correlations between different components suggests their shared contributions to $\Delta f^j_{\mathrm{eff}}, j = (0.1, 0.2, 0.5)$.

\begin{figure*}
\center
\includegraphics[width=0.97\textwidth]{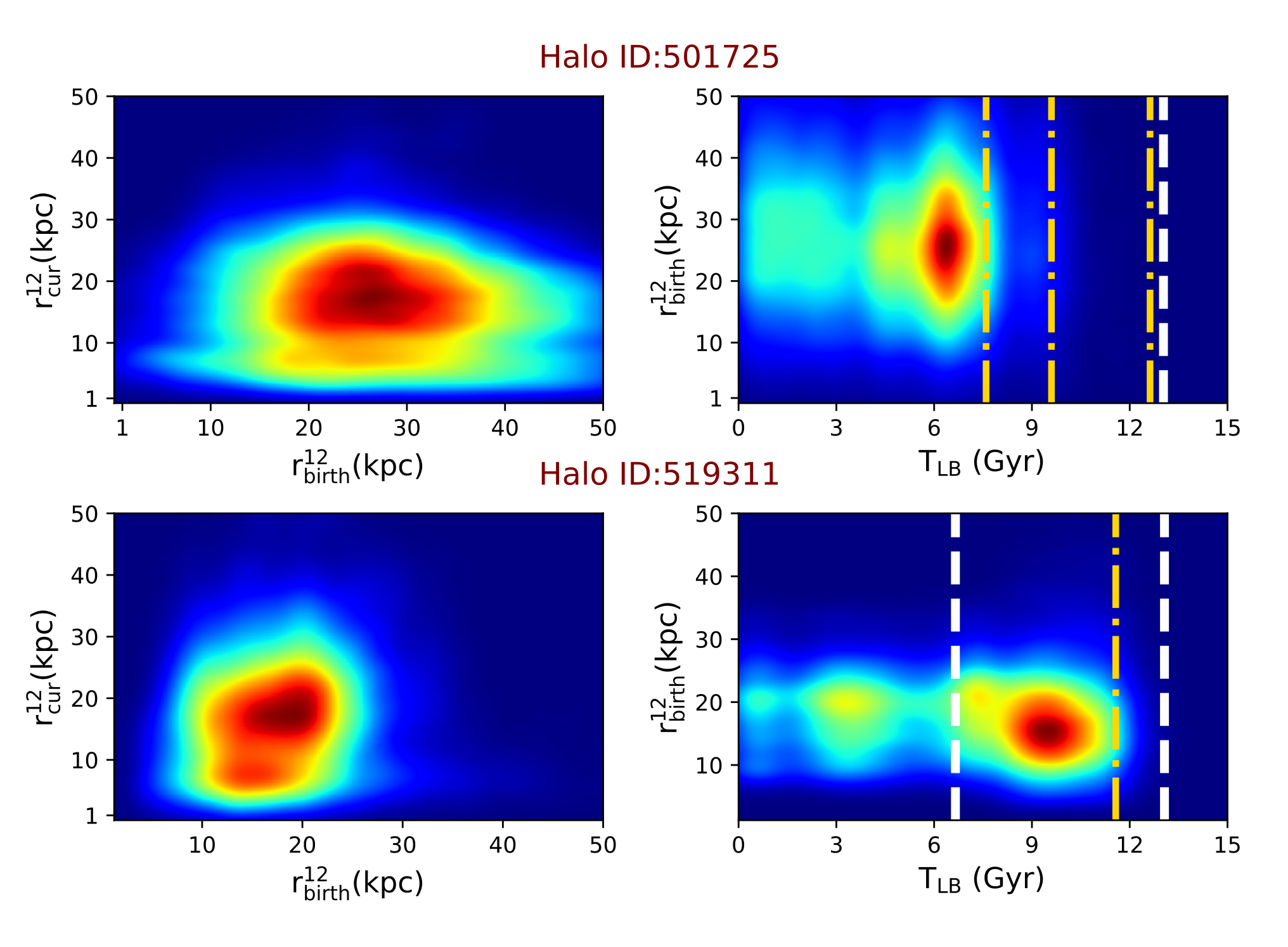}
\caption{Rows present the birth location vs the current location (left panels) as well as the look-back time (T$_{\mathrm{LB}}$) vs the birth location (right panels) of the in~situ stars located between the 0.1 R$_{\mathrm{vir}}$ and 0.2 R$_{\mathrm{vir}}$ in two galaxies of the sample analysed in this work. The dotted-dashed-gold and dashed-white lines refer to the minor and major mergers, respectively. Both minor and major mergers are important in elevating star formation and producing in~situ stars. In addition, in~situ stars get closer to the centre after their birth. }
\label{Rbirth-Rcurrent}
\end{figure*}

\subsection{External drivers for the in~situ fraction}
\label{External-drivers}
During the evolutionary trajectory of galaxies, numerous external events, including mergers and disruptions, play pivotal roles in shaping their dynamics. These events significantly influence star formation rates and redistribute stars within galaxies. Our objective here is to investigate how galaxy mergers impact the distribution of in~situ stars.  

Figure \ref{Rbirth-Rcurrent} illustrates the influence of galaxy mergers on the creation and distribution of in~situ stars in two galaxies (depicted in the top and bottom rows) of our galaxy sample. The left panels depict the birth location versus the current location, while the right panels display the look-back time (TLB) versus the birth location of the in~situ stars located between 0.1-0.2 $R_{\mathrm{vir}}$. The dotted-dashed-gold and dashed-white lines denote minor and major mergers, respectively. We have consistently computed the look-back time for all of the stars of interest as well as for mergers.
Focusing on the right panels of Figure \ref{Rbirth-Rcurrent}, it is evident that galaxy mergers contribute to the formation of high-density peaks in the stellar distribution, appearing as bright regions in the kernel density estimator (KDE). For instance, in the top (bottom) panel, a major merger occurring at $T_{\mathrm{LB}} \simeq $ 7.5 (11.5) Gyr results in the formation of a distinct density peak. Additionally, in the bottom panel, the KDE shape is noticeably altered following a minor merger at $T_{\mathrm{LB}} \simeq $ 6.5 Gyr, highlighting the impact of mergers on the stellar distribution.

Both minor and major mergers play a significant role in enhancing star formation and generating in~situ stars. Moreover, the analysis suggests that in~situ stars tend to migrate closer to the centre following their birth. Motivated by this, in the following, we compile a comprehensive list of potentially influential parameters and analyse their correlation coefficients with $\Delta f^j_{\mathrm{eff}}, j = (0.1, 0.2, 0.5)$. We provide detailed definitions of each parameter and elucidate their respective contributions to the observed deviations between the TNG50 and H3 datasets.

$\bullet$ \textbf{Effective mass-ratio:} One of the crucial parameters significantly influencing $\Delta f^j_{\mathrm{eff}}$, $j = (0.1, 0.2, 0.5)$, is the mass-ratio of galaxy mergers. Motivated by this insight, we introduce a metric termed the effective mass-ratio of galaxy mergers, $\overline{\mathrm{MR}}$, which encompasses both stellar mass and star-forming gas mass-ratios, where we define star-forming gas as cells with instantaneous star formation, identified by selecting cells with a positive ``StarFormationRate" in their gas field. While the role of the stellar mass-ratio, between first and second progenitors, is relatively well-understood, the influence of star-forming gas warrants further investigation. We posit that a gas-rich galaxy merger collectively enhances star formation. Consequently, we incorporate this additional contribution into our metric and observe a consistent correlation coefficient across all cases.

For each of these contributions, we compute the average mass-ratio above a specified threshold, denoted as $f_{\mathrm{MM}}$, after a designated redshift threshold. In subsequent analyses, we explore the impact of two values for the mass-ratio threshold, specifically $f_{\mathrm{MM}}=(0.05, 0.2)$. Additionally, through exploratory investigations, we determine that a redshift threshold of $z \leq 4$ effectively provides a better correlation, as the majority of galaxy evolution occurs below this threshold.  Furthermore, it is observed that this threshold is consistent with the mass-weighted mass-ratio as defined in Appendix \ref{mass-ratio-eff}. 
\begin{figure*}
\center
\includegraphics[width=0.88\textwidth]{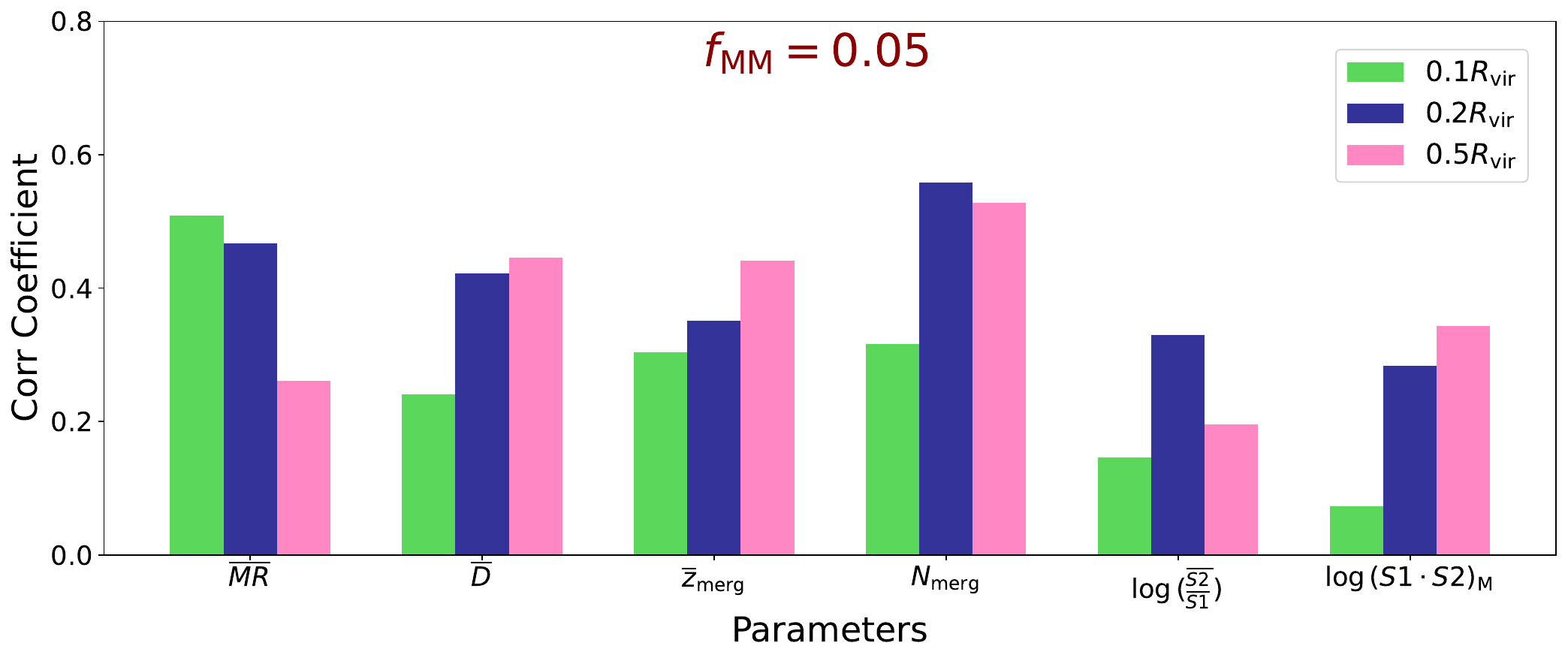}
\includegraphics[width=0.88\textwidth]{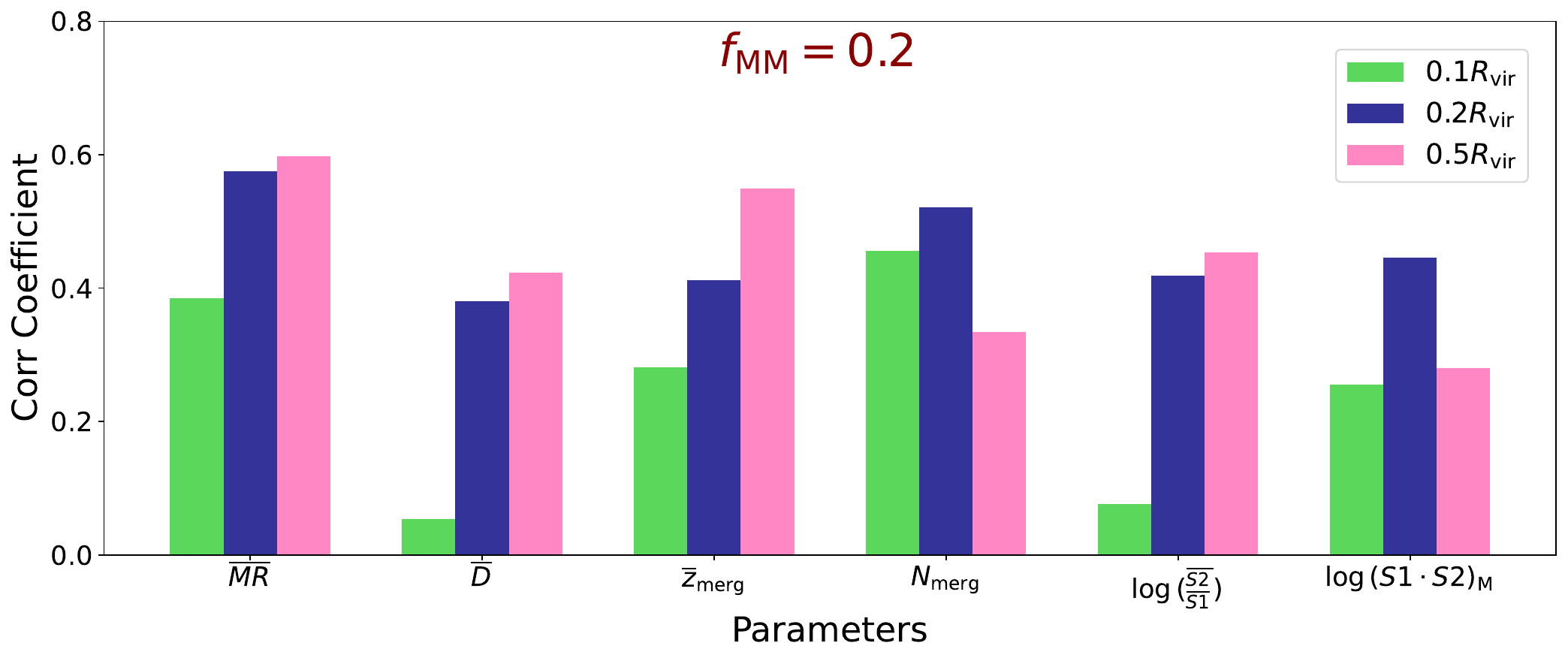}
\caption{The correlation coefficient between parameters associated with galaxy mergers and $\Delta f^j_{\mathrm{eff}}, j = (0.1, 0.2, 0.5)$.  The top row presents the correlation coefficient for the galaxy mergers above the effective mass-ratio 0.05, while the bottom row depicts the correlation coefficient for mergers with mass-ratio above 0.2. }
 \label{correlation-matrix-mass-ratio}
\end{figure*}
$\overline{\mathrm{MR}}$ is defined as:
\begin{equation}
\label{Mass-ratio}
\overline{\mathrm{MR}} = \frac{1}{N}\sum_{i} ^{N} \left(\mathrm{MR}_{*} + \mathrm{MR}_{g} \right)_i,
\end{equation}
where the summation index $i$ is over the total number of the mergers with the mass-ration above $f_{\mathrm{MM}}$. In each merger event, we identify the first and the second progenitors with 
$M^f_i$ and $M^s_i$, respectively. Furthermore, $\mathrm{MR}_{j} \equiv M^s_i/M^f_i$, $j = *$, and $j= g$ denote the mass-ratio of stars to star-forming gas, respectively. 

$\bullet$ \textbf{Mean distance:} The second crucial parameter, correlated with $\Delta f^j_{\mathrm{eff}}, j = (0.1, 0.2, 0.5)$, pertains to the impact parameter in galaxy mergers. In the event of a galaxy merger, the distance between the first and second galaxy significantly influences the redistribution of stars. Our analysis involves computing the average distance, hereafter referred to as $\overline{D}$, between the first and second galaxy of any galaxy mergers with mass-ratio above $f_{\mathrm{MM}}$ happening below the $z \leq 4$. The distance is inferred at the time when the second progenitor gets to the peak of its mass. Since we do not have enough  time resolution in the cosmological simulations, it is common to approximate this snapshot, or its subsequent snapshot, at the time of the merger. 

\begin{figure*}
\center
\includegraphics[width=0.47\textwidth]{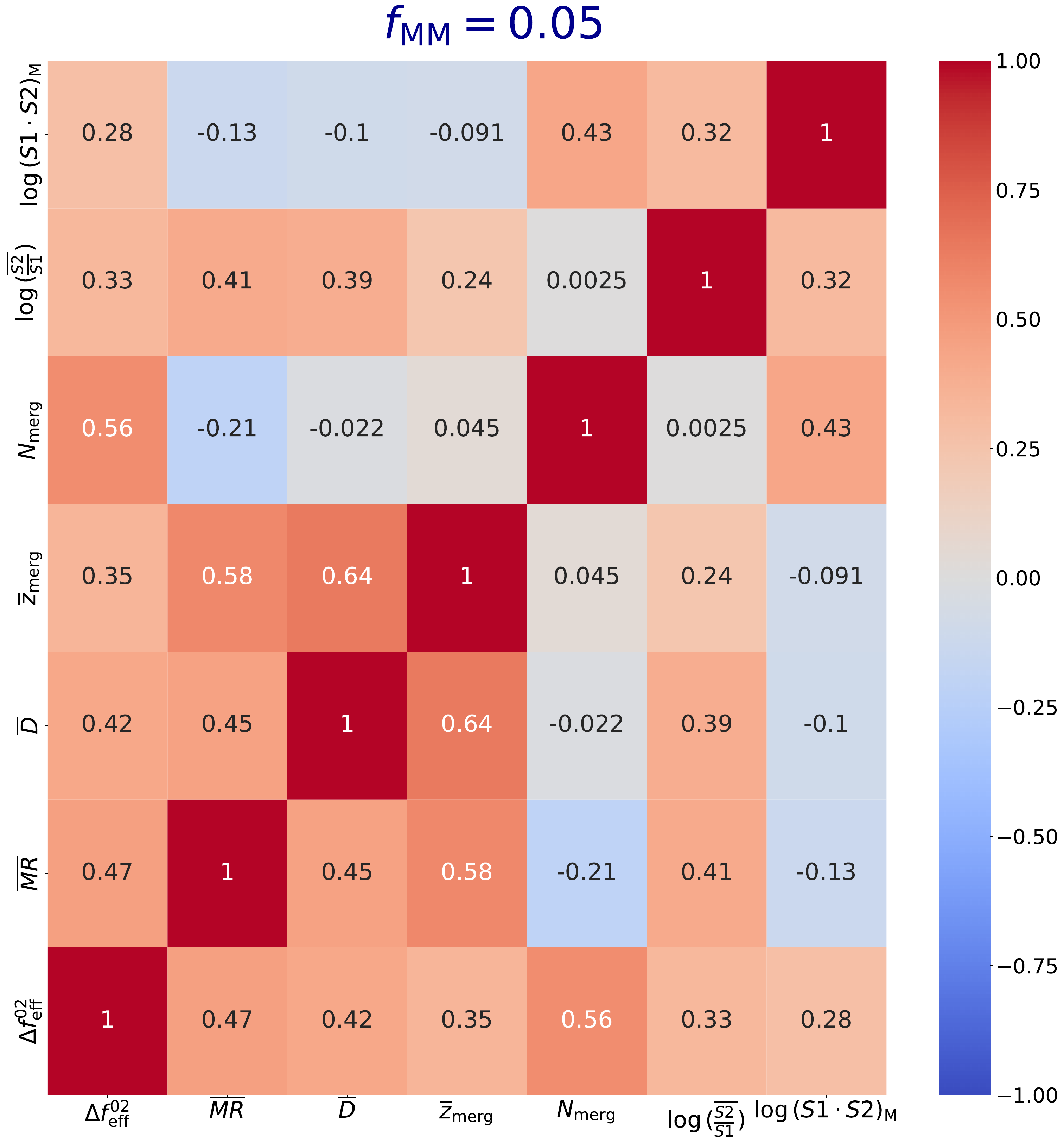}
\includegraphics[width=0.47\textwidth]{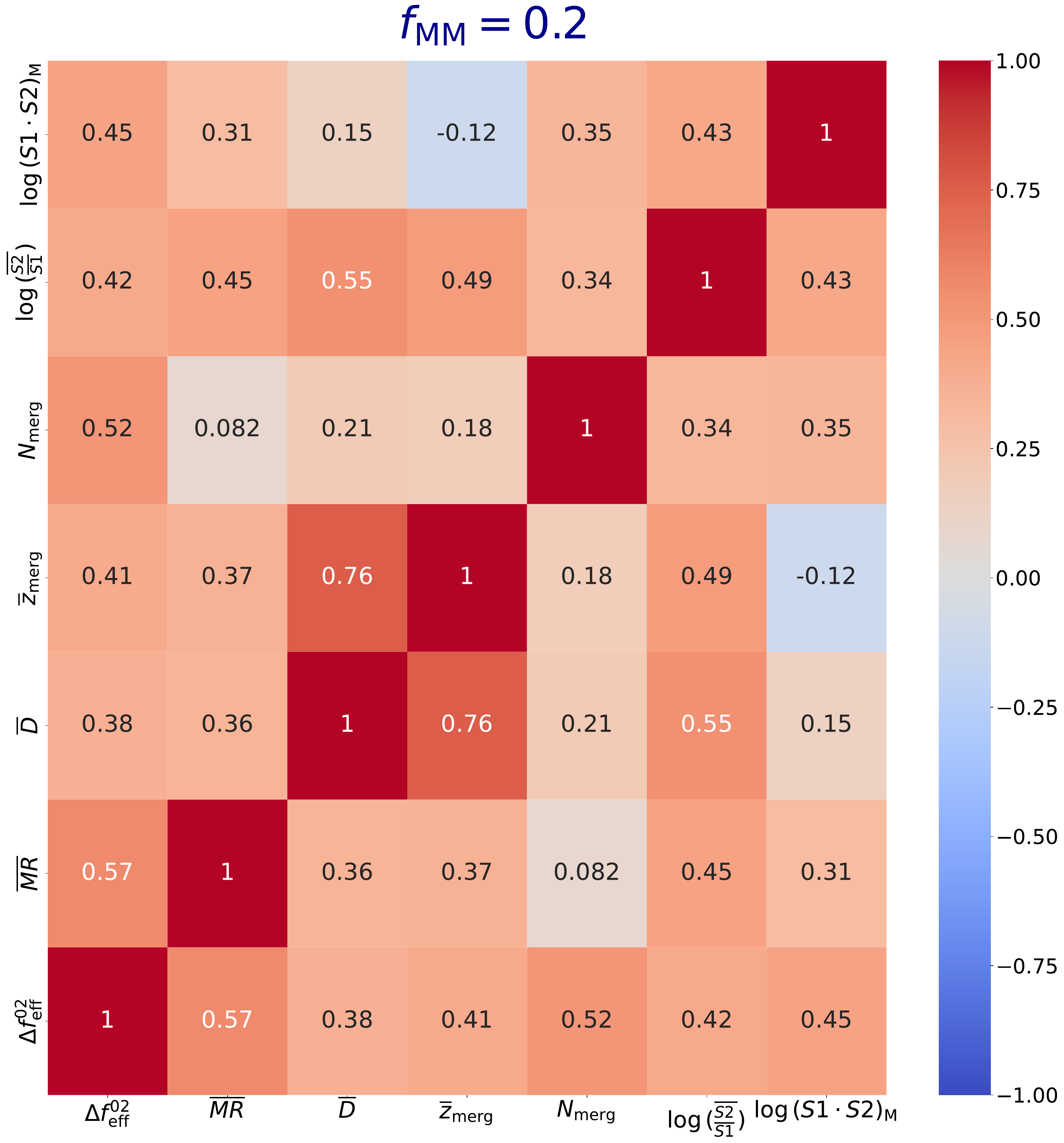}
\caption{Depiction of the correlation coefficient matrix illustrating the relationships between external parameters and $\Delta f^j_{\mathrm{eff}}$, as well as among themselves. The left panel corresponds to $f_{\mathrm{MM}}=0.05$, while the right panel illustrates the case with $f_{\mathrm{MM}}=0.2$.} 
\label{correlation-matrix}
\end{figure*}
$\bullet$ \textbf{Effective redshift:} The third critical parameter influencing $\Delta f^j_{\mathrm{eff}}, j = (0.1, 0.2, 0.5)$, is the timing of galaxy mergers. The occurrence of a galaxy merger during the primary epoch of galaxy evolution may exert a more pronounced effect on the stellar profile compared to mergers happening later. We establish an effective redshift of mergers, denoted as $z_{\mathrm{merg}}$, for mergers transpiring below $z \leq 4$ and surpassing a certain merger fraction threshold $f_{\mathrm{MM}}$. This effective redshift is calculated as $z_{\mathrm{merg}} = (1/N)\sum_{i}^{N} z_i$, where $N$ represents the total number of galaxy mergers, while $z_i$ describes the redshift associated with the $i$th merger event.

$\bullet$ \textbf{Number of mergers:} The fourth pivotal parameter is the total number of mergers with mass-ratios exceeding $f_{\mathrm{MM}}$ occurring below $z \leq 4$. We denote this quantity as $N_{\mathrm{merg}}$ and explore its correlation with $\Delta f^j_{\mathrm{eff}}, j = (0.1, 0.2, 0.5)$.
 
$\bullet$ \textbf{Mean spin ratio:} The fifth pivotal parameter associated with $\Delta f^j_{\mathrm{eff}}, j = (0.1, 0.2, 0.5)$, is the spin ratio between the first and second progenitors at the moment of a merger. Where the halo spin is defined as the total angular momentum vector, computed as the mass-weighted sum of the product of a particle’s coordinate and velocity for all particles or gas cells associated with the halo. From a physical perspective, in the event of a galaxy merger, both the spin magnitude and the spin alignment between the first and second progenitor play a crucial role. The spin magnitude, or more precisely the spin ratio, encodes how the stellar distribution is expected to evolve post-merger. A higher spin ratio between the second and first progenitor suggests that stars are more likely to be scattered farther after the merger, increasing the deviation between the in-situ fraction in TNG50 and the H3 survey. \\
Spin alignment is also expected to be a key factor, particularly in distinguishing between face-on and edge-on merger configurations.
We define the average value of this spin ratio, incorporating all galaxy mergers with mass-ratios exceeding $f_{\mathrm{MM}}$ occurring below $z \leq 4$ throughout the merger history of galaxies, and denote this quantity as $\overline{\left(\frac{S2}{S1}\right)}$. Our investigation revealed that taking the logarithm of this quantity yields a stronger correlation. Therefore, we employ $\log{(\overline{\frac{S2}{S1}})}$ when computing the correlation coefficients. 

$\bullet$ \textbf{ Maximum spin alignment:} The final critical parameter associated with $\Delta f^j_{\mathrm{eff}}, j = (0.1, 0.2, 0.5)$, is the spin alignment between the first and second progenitors. We differentiate between the impact of face-on and edge-on galaxy mergers on the redistribution of stars across galaxies. Our investigation has demonstrated that the maximum alignment, derived from a subset of galaxy mergers meeting the aforementioned criteria, yields a stronger correlation coefficient. Furthermore, it has been observed that the logarithm of this quantity provides an even higher correlation. We define this parameter as $\log{(S1 \cdot S2)_{\mathrm{M}}}$.

\subsubsection{Corr(External driver, $\Delta f^j_{\mathrm{eff}}$)}
\label{X-deltaf-correlation}
Figure \ref{correlation-matrix-mass-ratio} illustrates the correlation function between individual external parameters and $\Delta f^j_{\mathrm{eff}}$. The first and second rows present the correlation coefficients for $f_{\mathrm{MM}} = 0.05$ and $f_{\mathrm{MM}} = 0.2$, respectively. Within each row, different colours represent distinct spatial cuts: green, blue, and pink correspond to 0.1$R_{\mathrm{vir}}$, 0.2$R_{\mathrm{vir}}$, and 0.5$R_{\mathrm{vir}}$, respectively.

From the diagram, it is evident that all external drivers are positively correlated with $\Delta f^j_{\mathrm{eff}}$, although the strength of correlation coefficients varies across different spatial cuts and mass-ratio merger thresholds. It is anticipated that higher values of $\overline{\mathrm{MR}}$ and $N_{\mathrm{merg}}$ result in larger $\Delta f^j_{\mathrm{eff}}$. 
It is also seen that $\overline{D}$ positively correlates with $\Delta f^j_{\mathrm{eff}}$. We argue that mergers with larger $\overline{D}$, unless those with very close to equal mass ratios, will involve galaxies with multiple encounters with the first progenitor before the actual merger. This may trigger both star formation as well as further enhance the scattering in the stellar population. Furthermore, we speculate that galaxy mergers with larger impact parameters may also leave star-forming gas at larger distances, causing them to form stars at larger radii.\\ 
$z_{\mathrm{merg}}$  is also positively correlated with $\Delta f^j_{\mathrm{eff}}$. Our analysis suggests that having galaxy mergers at higher redshifts, albeit below $z=4$, is crucial for transporting star-forming gas to the outer regions of galaxies, thus facilitating star formation at subsequent redshifts. 

The highest correlation coefficient at $f_{\mathrm{MM}} = 0.05$ is attributed to $N_{\mathrm{merg}}$, suggesting its strongest association with $\Delta f^j_{\mathrm{eff}}$. Nevertheless, both $\overline{\mathrm{MR}}$ and $z_{\mathrm{merg}}$ also exhibit significant importance.

At $f_{\mathrm{MM}} = 0.05$, the lowest correlation coefficient is linked to the spin-driven quantities, although this changes at $f_{\mathrm{MM}} = 0.2$. At this level, the correlation coefficients for all quantities at 0.2$R_{\mathrm{vir}}$ increase, while some of them decrease for 0.1$R_{\mathrm{vir}}$. 
As shown in Figure \ref{correlation-matrix-mass-ratio}, more aligned mergers, where the progenitor spins are nearly parallel (i.e., face-on mergers), exhibit a higher correlation coefficient with $\Delta f^j_{\mathrm{eff}}, j = (0.1, 0.2, 0.5)$.
A deeper investigation with higher time resolution would be valuable to quantitatively assess the role of face-on vs. edge-on mergers in shaping this correlation. However, such an analysis is beyond the scope of this study and is left for future work.

While Figure \ref{correlation-matrix-mass-ratio} illustrates the significance of individual parameters in influencing $\Delta f^j_{\mathrm{eff}}$, it does not quantify the extent to which these external drivers may overlap or exhibit degeneracy in their correlation coefficients. To quantify potential levels of degeneracy between different external drivers, we proceed by computing the cross-correlation between these quantities.

\subsubsection{Corr(External driver, External driver)}
\label{X-X-correlation}
Figure \ref{correlation-matrix} depicts the correlation coefficient matrix, illustrating the relationships between external parameters and $\Delta f^j_{\mathrm{eff}}$, as well as among themselves. The left panel corresponds to $f_{\mathrm{MM}}=0.05$, while the right panel illustrates the case with $f_{\mathrm{MM}}=0.2$.

Positive and negative cross-correlations are reported, although the statistical significance of negative correlations is notably smaller than that of positive cases. Hence, our focus lies solely on positive correlations. Notably, $\overline{\mathrm{MR}}$ exhibits strong correlations with $\overline{D}$, $z_{\mathrm{merg}}$, and $\log{(\overline{\frac{S2}{S1}})}$. This is attributed to higher mass-ratios potentially exerting stronger impacts on average impact parameter values. Additionally, the positive correlation with $z_{\mathrm{merg}}$ suggests that mergers with higher mass-ratios may tend to occur at higher redshifts. Furthermore, a positive correlation with $\log{(\overline{\frac{S2}{S1}})}$ indicates that galaxy mergers with higher mass-ratios may also correspond to higher spin fractions, although alignment is not necessarily guaranteed.

$\overline{D}$ is correlated with $z_{\mathrm{merg}}$, hinting that galaxy mergers at higher redshifts may involve larger impact parameters. While this observation is intriguing, further justification will be pursued in future work with a larger galaxy sample.

Additionally, the number of mergers shows a positive correlation with $\log{(S1 \cdot S2)_{\mathrm{M}}}$, suggesting that a larger number of galaxy mergers statistically facilitates the occurrence of more aligned systems. Moreover, there are positive correlations between $\log{(\overline{\frac{S2}{S1}})}$ and $\log{(S1 \cdot S2)_{\mathrm{M}}}$, indicating that, on average, systems with higher spin fractions may exhibit higher alignments.

These observations reveal overlapping effects between key drivers, ultimately enhancing their correlation coefficients with $\Delta f^j_{\mathrm{eff}}$. Consequently, it is imperative to be cautious when listing the key external parameters, as their impacts may be lower due to the aforementioned degeneracy with other parameters.

\section{Conclusions}
\label{conclusion}

In this study, we conducted a comprehensive analysis of the distribution of in~situ stars from a sample of Milky Way-like galaxies from the TNG50 simulation. The identification process relied solely on the stellar birth location within the first progenitor throughout the galaxy's evolution. We investigated the effects of various spatial cuts, including 30 kpc, 0.1$R_{\mathrm{vir}}$, 0.2$R_{\mathrm{vir}}$, and 0.5$R_{\mathrm{vir}}$, on the profile distribution of in~situ stars, along with an additional spatial cut akin to the heliocentric coordinate of the H3 survey. Subsequently, we compared the scale height distribution of in~situ stars from TNG50 with the latest observational outcomes from the H3 survey, identifying the impact of different internal and external parameters in inducing deviations between theory and observation.

Below, we summarise some of the key steps and highlight a few takeaways from this exploratory investigation.

1. We conducted an analysis to determine the correlation coefficient between $\Delta f^j_{\mathrm{eff}}$ from Equation \ref{Effective-Rank} and internal drivers, such as the mean/median of the amplitude and derivatives $R_{\mathrm{vir}}$ and the SFR over various time intervals. These intervals encompassed either the period from an initial redshift to $z_{\mathrm{cut}}$ or from $z_{\mathrm{cut}}$ to redshift zero.

2. The observed positive correlation between $\widetilde{|\frac{dR_{\mathrm{vir}}}{dz}|}_H$ and $\Delta f^j_{\mathrm{eff}}$
in Figure \ref{correlation-matrix-internal}, particularly at higher $z_{\mathrm{cut}}$ values, indicates that the expansion of $R_{\mathrm{vir}}$ encompasses more in~situ stars, leading to an increase in $\Delta f^j_{\mathrm{eff}}$.

3. At lower $z_{\mathrm{cut}}$ values, Figure \ref{correlation-matrix-internal} implies that $\widetilde{|\frac{dR_{\mathrm{vir}}}{dz}|}_L$ inversely correlates with $\Delta f^j_{\mathrm{eff}}$, indicating a contraction of spatial boundaries with decreasing $R_{\mathrm{vir}}$ and lower $\Delta f^j_{\mathrm{eff}}$ values. Conversely, at higher $z_{\mathrm{cut}}$, this correlation shifts to a positive correlation, possibly due to changes in $R_{\mathrm{vir}}$ slope with increasing $z_{\mathrm{cut}}$.

4. Figure \ref{correlation-matrix-internal} illustrates varying correlations between $\overline{\mathrm{SFR}}_L$ and $\Delta f^j_{\mathrm{eff}}$ across different $z_{\mathrm{cut}}$ values. Stronger correlations are evident at smaller spatial cuts for lower $z_{\mathrm{cut}}$, whereas they occur at larger spatial cuts for higher $z_{\mathrm{cut}}$. 

5. The non-zero cross-correlation observed among different internal drivers, as depicted in Figure \ref{Cross-Correlation-internal}, indicates their shared influence on $\Delta f^j_{\mathrm{eff}}$.

6. Our analysis, as depicted in Figure \ref{Rbirth-Rcurrent}, emphasises the considerable impact of mergers on star formation,  being evident as the appearance of density peaks in the KDE after a merger occurs. Additionally, it reveals a consistent pattern of in~situ stars migrating towards the galactic centre post-formation.

7. Moreover, our analysis revealed six key parameters linked to galaxy mergers that notably influence the creation and distribution of in~situ stars. These parameters include the merger effective mass-ratio, mean distance between galaxy mergers, effective redshift of mergers, number of mergers, mean spin ratio of galaxy mergers, and maximum spin alignments from galaxy mergers.

8. The correlation analysis, presented in Figure \ref{correlation-matrix-mass-ratio}, reveals positive correlations between all external drivers and $\Delta f^j_{\mathrm{eff}}$, albeit with varying strengths across different spatial cuts and mass-ratio merger thresholds. Higher values of $\overline{\mathrm{MR}}$ and $N_{\mathrm{merg}}$ are expected to increase $\Delta f^j_{\mathrm{eff}}$. 
Our analysis suggested that $\overline{D}$ positively correlates with $\Delta f^j_{\mathrm{eff}}$. We argue that mergers with larger $\overline{D}$, unless those with very close to equal mass ratios, will involve galaxies with multiple encounters with the first progenitor before the actual merger. This may trigger both star formation as well as further enhance the scattering in the stellar population.

9. Our analysis, illustrated in Figure \ref{correlation-matrix}, reveals positive cross-correlations between different external parameters, suggesting some degree of degeneracy in influencing $\Delta f^j_{\mathrm{eff}}$. Notably, a strong correlation is observed between $\overline{\mathrm{MR}}$ and $\overline{D}$, $z_{\mathrm{merg}}$, and $\log{(\overline{\frac{S2}{S1}})}$. Additionally, $\overline{D}$ is correlated with $z_{\mathrm{merg}}$. Lastly, $N_{\mathrm{merg}}$ exhibits a positive correlation with $\log{(S1 \cdot S2)_{\mathrm{M}}}$, while $\log{(S1 \cdot S2)_{\mathrm{M}}}$ is also correlated with 
$\log{(S1 \cdot S2)_{\mathrm{M}}}$. 

10. While the significant correlations observed between external parameters and $\Delta f^j_{\mathrm{eff}}$, as well as among themselves, are intriguing, it is important to note that the statistical reliability of some correlation coefficients may be impacted by the small sample size. To address this concern, we employed a Random Forest Regression method in Appendix \ref{Random-Forest-Regression} to assess the generalizability of these results using a machine learning approach. The results, depicted in Figure \ref{correlation-matrix-comparison}, indicate that the correlation strengths between these parameters and $\Delta f^j_{\mathrm{eff}}$ remain relatively consistent after conducting the Random Forest Regression.

\section*{Future Directions}
In future studies, we aim to expand our galaxy sample size to a bigger sample to further validate our conclusions. Additionally, we plan to investigate the influence of mergers on individual galaxy spins and star formation by utilising isolated galaxy merger simulations and zoom-in simulations with increased spatial and temporal resolution. Finally, while in the current study, we mainly followed a theoretically driven approach to identify the in~situ stars based on a spatial cut, in a follow-up work we plan to extend this fundamental study by taking a more chemically oriented approach in defining the in~situ stars. While our initial investigations demonstrated that the results are not heavily sensitive to this choice, it is worth checking how much this conclusion might be changed if we use an alternative cosmological simulation with a better-defined stellar population. While our analysis focuses on a spatial cut based on the virial radius, an alternative approach would be to investigate the impact of selecting stars based on gravitational potential. However, this lies beyond the scope of the present study and is left for future work.

\section*{Data Availability}
Data directly associated with this manuscript and its figures can be provided upon reasonable request from the corresponding author. The IllustrisTNG and TNG50 simulations are publicly available and accessible at \url{www.tng-project.org/data} \citep{2019ComAC...6....2N}.

\section*{acknowledgement}
It is a great pleasure to warmly acknowledge Charlie Conroy, Daniel Eisenstein, Rohan Naidu, Ramesh Narayan, Matthew Liska, Sirio Belli, Sandro Tacchella, Kaylee Desoto, and Charles Alcock for very fruitful conversations and constructive comments during this work. We express our deep gratitude to the referee for their constructive comments, which have significantly improved the quality of this paper.
RE acknowledges the support of the Institute for Theory and Computation at the Center for Astrophysics as well as grant numbers 21-atp21-0077, and HST-GO-16173.001-A. SB is supported by the UK Research and Innovation (UKRI) Future Leaders Fellowship (grant number MR/V023381/1). We thank the useful conversation at the AstroAI Institute at the CfA. We thank the supercomputer facility at Harvard where most of the simulation work was done. MV acknowledges support through an MIT RSC award, a Kavli Research Investment Fund, NASA ATP grant NNX17AG29G, and NSF grants AST-1814053, AST-1814259, and AST-1909831. The TNG50 simulation was realised with compute time granted by the Gauss Centre for Supercomputing (GCS) under GCS Large-Scale Projects GCS-DWAR on the GCS share of the supercomputer Hazel Hen at the High-Performance Computing Center Stuttgart (HLRS). 

\textit{Software:} matplotlib \citep{2007CSE.....9...90H}, numpy \citep{2011CSE....13b..22V}, scipy \citep{2007CSE.....9c..10O}, seaborn \citep{2020zndo...3629446W}, pandas \citep{2020zndo...3629446W}, h5py \citep{2016arXiv160804904D}.

\appendix

\section{Effective mass-ratio}
\label{mass-ratio-eff}

As already stated in Equation \ref{Mass-ratio}, the mass-ratio is defined by combining the stellar mass-ratio and star-forming gas ratio. While this quantity is closely connected to the other external parameters, as well as $\Delta f^j_{\mathrm{eff}}$, we chose a redshift threshold of $z \leq 4$ for including mergers in the system. This threshold was chosen to ensure that we consider enough prior galaxy evolution while also incorporating the main evolution of galaxies into account. To verify the justification of this threshold, we extend beyond the former definition and define a mass-weighted merger mass-ratio without requiring a specific redshift threshold. 
More explicitly, for each merger event, in the numerator, we add an extra factor of the mass of the second progenitor. This is given as: 
\begin{equation}
\label{Effective-MR}
\mathrm{MR}_{\mathrm{eff}} \equiv  \frac{\sum_{i=1}^{N} \left(\frac{\left(M^s_{*}\right)^2}{M^f_{*}} \right)_i} {\sum_{i=1}^{N} (M^s_{*}) _i} + \frac{\sum_{i=1}^{N} \left(\frac{\left(M^s_{g}\right)^2}{M^f_{g}} \right)_i} {\sum_{i=1}^{N} (M^s_{g}) _i}.
\end{equation}
where $M^j_{*}, j = (f,s)$ refers to the stellar mass in the first and second progenitors, while $M^j_{g}, j = (f,s)$ describes the gaseous mass in the first/second progenitor, respectively. 
Our analysis demonstrated that the correlation level between this new mass-ratio and $\Delta f^j_{\mathrm{eff}}$ is compatible with that of $\overline{\mathrm{MR}}$. Consequently, we focus exclusively on using the former quantity throughout our analysis. 
 
\begin{figure*}
\center
\includegraphics[width=0.88\textwidth]{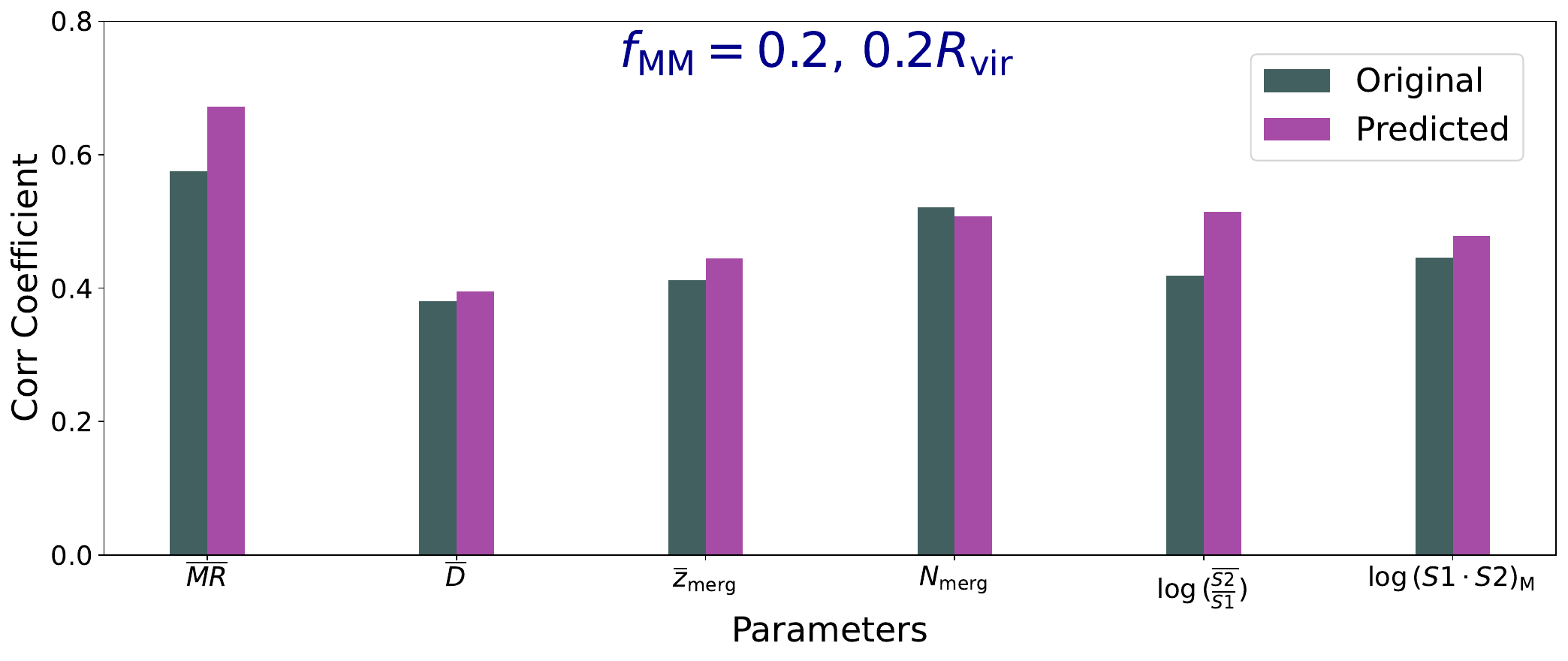}
\caption{Comparison between the original correlation matrix and the predicted one using the Random Forest Regression for external drivers for the major merger mass-ratio above 0.2 and at $0.2 R_{\mathrm{vir}}$ spatial cut. The fit corresponds to a $R^2 = 0.9$.}
 \label{correlation-matrix-comparison}
\end{figure*}


\section{Generalisation to bigger sample: Random Forest Regression}
\label{Random-Forest-Regression}
In the main body of the paper, we analysed the correlation coefficients between different internal and external parameters with $\Delta f^j_{\mathrm{eff}}$. While the level of correlation is significant for the parameter sets used, it is important to exercise caution regarding the generalizability of this analysis, given that our galaxy sample only includes 25 members.

To address the generalizability of the correlation coefficients between the external drivers and $\Delta f^{02}_{\mathrm{eff}}$, we employed Random Forest Regression to infer the predicted correlation coefficients between these quantities. Since the sample size is limited, instead of splitting the sample into training and test datasets, we  used k-fold cross-validation. Our analysis indicated that using a k-fold value of 3 yielded a reasonable $R^2 \simeq 0.34$, where $R^2$ represents the proportion of the variance in the dependent variable that is explained by the independent variables in the regression model. Consequently, we employed Random Forest Regression to infer the correlation coefficients.

Figure \ref{correlation-matrix-comparison} compares the correlation coefficients from the original galaxy sample with those predicted by Random Forest Regression. The level of correlation is consistent between these methods. Furthermore, the model reported a reliable value for $R^2 \simeq 0.9$, ensuring that these correlation coefficients can be representative of a larger system.

\newpage 

\bibliography{main}{}

\begin{thebibliography}{}
\expandafter\ifx\csname natexlab\endcsname\relax\def\natexlab#1{#1}\fi
\providecommand{\url}[1]{\href{#1}{#1}}
\providecommand{\dodoi}[1]{doi:~\href{http://doi.org/#1}{\nolinkurl{#1}}}
\providecommand{\doeprint}[1]{\href{http://ascl.net/#1}{\nolinkurl{http://ascl.net/#1}}}
\providecommand{\doarXiv}[1]{\href{https://arxiv.org/abs/#1}{\nolinkurl{https://arxiv.org/abs/#1}}}

\bibitem[{{Bell} {et~al.}(2008){Bell}, {Zucker}, {Belokurov}, {Sharma},
  {Johnston}, {Bullock}, {Hogg}, {Jahnke}, {de Jong}, {Beers}, {Evans},
  {Grebel}, {Ivezi{\'c}}, {Koposov}, {Rix}, {Schneider}, {Steinmetz}, \&
  {Zolotov}}]{2008ApJ...680..295B}
{Bell}, E.~F., {Zucker}, D.~B., {Belokurov}, V., {et~al.} 2008, \apj, 680, 295,
  \dodoi{10.1086/588032}

\bibitem[{{Belokurov} {et~al.}(2020){Belokurov}, {Sanders}, {Fattahi}, {Smith},
  {Deason}, {Evans}, \& {Grand}}]{2020MNRAS.494.3880B}
{Belokurov}, V., {Sanders}, J.~L., {Fattahi}, A., {et~al.} 2020, \mnras, 494,
  3880, \dodoi{10.1093/mnras/staa876}

\bibitem[{{Benson} {et~al.}(2004){Benson}, {Lacey}, {Frenk}, {Baugh}, \&
  {Cole}}]{2004MNRAS.351.1215B}
{Benson}, A.~J., {Lacey}, C.~G., {Frenk}, C.~S., {Baugh}, C.~M., \& {Cole}, S.
  2004, \mnras, 351, 1215, \dodoi{10.1111/j.1365-2966.2004.07870.x}

\bibitem[{{Bonaca} {et~al.}(2017){Bonaca}, {Conroy}, {Wetzel}, {Hopkins}, \&
  {Kere{\v{s}}}}]{2017ApJ...845..101B}
{Bonaca}, A., {Conroy}, C., {Wetzel}, A., {Hopkins}, P.~F., \& {Kere{\v{s}}},
  D. 2017, \apj, 845, 101, \dodoi{10.3847/1538-4357/aa7d0c}

\bibitem[{{Carollo} \& {Chiba}(2021)}]{2021ApJ...908..191C}
{Carollo}, D., \& {Chiba}, M. 2021, \apj, 908, 191,
  \dodoi{10.3847/1538-4357/abd7a4}

\bibitem[{{Conroy} {et~al.}(2019{\natexlab{a}}){Conroy}, {Naidu}, {Zaritsky},
  {Bonaca}, {Cargile}, {Johnson}, \& {Caldwell}}]{2019ApJ...887..237C}
{Conroy}, C., {Naidu}, R.~P., {Zaritsky}, D., {et~al.} 2019{\natexlab{a}},
  \apj, 887, 237, \dodoi{10.3847/1538-4357/ab5710}

\bibitem[{{Conroy} {et~al.}(2019{\natexlab{b}}){Conroy}, {Bonaca}, {Cargile},
  {Johnson}, {Caldwell}, {Naidu}, {Zaritsky}, {Fabricant}, {Moran}, {Rhee},
  {Szentgyorgyi}, {Berlind}, {Calkins}, {Kattner}, \&
  {Ly}}]{2019ApJ...883..107C}
{Conroy}, C., {Bonaca}, A., {Cargile}, P., {et~al.} 2019{\natexlab{b}}, \apj,
  883, 107, \dodoi{10.3847/1538-4357/ab38b8}

\bibitem[{{Cooper} {et~al.}(2015){Cooper}, {Parry}, {Lowing}, {Cole}, \&
  {Frenk}}]{2015MNRAS.454.3185C}
{Cooper}, A.~P., {Parry}, O.~H., {Lowing}, B., {Cole}, S., \& {Frenk}, C. 2015,
  \mnras, 454, 3185, \dodoi{10.1093/mnras/stv2057}

\bibitem[{{Cui} {et~al.}(2012){Cui}, {Zhao}, {Chu}, {Li}, {Li}, {Zhang}, {Su},
  {Yao}, {Wang}, {Xing}, {Li}, {Zhu}, {Wang}, {Gu}, {Luo}, {Xu}, {Zhang},
  {Liu}, {Zhang}, {Yang}, {Cao}, {Chen}, {Chen}, {Chen}, {Chen}, {Chu}, {Feng},
  {Gong}, {Hou}, {Hu}, {Hu}, {Hu}, {Jia}, {Jiang}, {Jiang}, {Jiang}, {Jin},
  {Li}, {Li}, {Li}, {Liu}, {Liu}, {Lu}, {Mao}, {Men}, {Qi}, {Qi}, {Shi},
  {Tang}, {Tao}, {Wang}, {Wang}, {Wang}, {Wang}, {Wang}, {Wang}, {Wang},
  {Wang}, {Wang}, {Wang}, {Wang}, {Wang}, {Xu}, {Xu}, {Yang}, {Yu}, {Yuan},
  {Yuan}, {Zhai}, {Zhang}, {Zhang}, {Zhang}, {Zhao}, {Zhou}, {Zhou}, {Zhu}, \&
  {Zou}}]{2012RAA....12.1197C}
{Cui}, X.-Q., {Zhao}, Y.-H., {Chu}, Y.-Q., {et~al.} 2012, Research in Astronomy
  and Astrophysics, 12, 1197, \dodoi{10.1088/1674-4527/12/9/003}

\bibitem[{{de Buyl} {et~al.}(2016){de Buyl}, {Huang}, \&
  {Deprez}}]{2016arXiv160804904D}
{de Buyl}, P., {Huang}, M.-J., \& {Deprez}, L. 2016, arXiv e-prints,
  arXiv:1608.04904.
\newblock \doarXiv{1608.04904}

\bibitem[{{De Lucia} \& {Blaizot}(2007)}]{2007MNRAS.375....2D}
{De Lucia}, G., \& {Blaizot}, J. 2007, \mnras, 375, 2,
  \dodoi{10.1111/j.1365-2966.2006.11287.x}

\bibitem[{{De Silva} {et~al.}(2015){De Silva}, {Freeman}, {Bland-Hawthorn},
  {Martell}, {de Boer}, {Asplund}, {Keller}, {Sharma}, {Zucker}, {Zwitter},
  {Anguiano}, {Bacigalupo}, {Bayliss}, {Beavis}, {Bergemann}, {Campbell},
  {Cannon}, {Carollo}, {Casagrande}, {Casey}, {Da Costa}, {D'Orazi}, {Dotter},
  {Duong}, {Heger}, {Ireland}, {Kafle}, {Kos}, {Lattanzio}, {Lewis}, {Lin},
  {Lind}, {Munari}, {Nataf}, {O'Toole}, {Parker}, {Reid}, {Schlesinger},
  {Sheinis}, {Simpson}, {Stello}, {Ting}, {Traven}, {Watson}, {Wittenmyer},
  {Yong}, \& {{\v{Z}}erjal}}]{2015MNRAS.449.2604D}
{De Silva}, G.~M., {Freeman}, K.~C., {Bland-Hawthorn}, J., {et~al.} 2015,
  \mnras, 449, 2604, \dodoi{10.1093/mnras/stv327}

\bibitem[{{Di Matteo} {et~al.}(2019){Di Matteo}, {Haywood}, {Lehnert}, {Katz},
  {Khoperskov}, {Snaith}, {G{\'o}mez}, \& {Robichon}}]{2019A&A...632A...4D}
{Di Matteo}, P., {Haywood}, M., {Lehnert}, M.~D., {et~al.} 2019, \aap, 632, A4,
  \dodoi{10.1051/0004-6361/201834929}

\bibitem[{{Di Matteo} {et~al.}(2020){Di Matteo}, {Spite}, {Haywood},
  {Bonifacio}, {G{\'o}mez}, {Spite}, \& {Caffau}}]{2020A&A...636A.115D}
{Di Matteo}, P., {Spite}, M., {Haywood}, M., {et~al.} 2020, \aap, 636, A115,
  \dodoi{10.1051/0004-6361/201937016}

\bibitem[{{Eggen} {et~al.}(1962){Eggen}, {Lynden-Bell}, \&
  {Sandage}}]{1962ApJ...136..748E}
{Eggen}, O.~J., {Lynden-Bell}, D., \& {Sandage}, A.~R. 1962, \apj, 136, 748,
  \dodoi{10.1086/147433}

\bibitem[{{Elias} {et~al.}(2018){Elias}, {Sales}, {Creasey}, {Cooper},
  {Bullock}, {Rich}, \& {Hernquist}}]{2018MNRAS.479.4004E}
{Elias}, L.~M., {Sales}, L.~V., {Creasey}, P., {et~al.} 2018, \mnras, 479,
  4004, \dodoi{10.1093/mnras/sty1718}

\bibitem[{{Emami} {et~al.}(2020{\natexlab{a}}){Emami}, {Genel}, {Hernquist},
  {Alcock}, {Bose}, {Weinberger}, {Vogelsberger}, {Marinacci}, {Loeb},
  {Torrey}, \& {Forbes}}]{2020arXiv200909220E}
{Emami}, R., {Genel}, S., {Hernquist}, L., {et~al.} 2020{\natexlab{a}}, arXiv
  e-prints, arXiv:2009.09220.
\newblock \doarXiv{2009.09220}

\bibitem[{{Emami} {et~al.}(2020{\natexlab{b}}){Emami}, {Hernquist}, {Alcock},
  {Genel}, {Bose}, {Weinberger}, {Vogelsberger}, {Shen}, {Speagle},
  {Marinacci}, {Forbes}, \& {Torrey}}]{2020arXiv201212284E}
{Emami}, R., {Hernquist}, L., {Alcock}, C., {et~al.} 2020{\natexlab{b}}, arXiv
  e-prints, arXiv:2012.12284.
\newblock \doarXiv{2012.12284}

\bibitem[{{Emami} {et~al.}(2022){Emami}, {Hernquist}, {Vogelsberger}, {Shen},
  {Speagle}, {Moreno}, {Alcock}, {Genel}, {Forbes}, {Marinacci}, \&
  {Torrey}}]{2022ApJ...937...20E}
{Emami}, R., {Hernquist}, L., {Vogelsberger}, M., {et~al.} 2022, \apj, 937, 20,
  \dodoi{10.3847/1538-4357/ac86c7}

\bibitem[{{Fattahi} {et~al.}(2020){Fattahi}, {Deason}, {Frenk}, {Simpson},
  {G{\'o}mez}, {Grand}, {Monachesi}, {Marinacci}, \&
  {Pakmor}}]{2020MNRAS.497.4459F}
{Fattahi}, A., {Deason}, A.~J., {Frenk}, C.~S., {et~al.} 2020, \mnras, 497,
  4459, \dodoi{10.1093/mnras/staa2221}

\bibitem[{{Font} {et~al.}(2011){Font}, {McCarthy}, {Crain}, {Theuns}, {Schaye},
  {Wiersma}, \& {Dalla Vecchia}}]{2011MNRAS.416.2802F}
{Font}, A.~S., {McCarthy}, I.~G., {Crain}, R.~A., {et~al.} 2011, \mnras, 416,
  2802, \dodoi{10.1111/j.1365-2966.2011.19227.x}

\bibitem[{{Gaia Collaboration} {et~al.}(2016){Gaia Collaboration}, {Prusti},
  {de Bruijne}, {Brown}, {Vallenari}, {Babusiaux}, {Bailer-Jones}, {Bastian},
  {Biermann}, {Evans}, {Eyer}, {Jansen}, {Jordi}, {Klioner}, {Lammers},
  {Lindegren}, {Luri}, {Mignard}, {Milligan}, {Panem}, {Poinsignon},
  {Pourbaix}, {Randich}, {Sarri}, {Sartoretti}, {Siddiqui}, {Soubiran},
  {Valette}, {van Leeuwen}, {Walton}, {Aerts}, {Arenou}, {Cropper}, {Drimmel},
  {H{\o}g}, {Katz}, {Lattanzi}, {O'Mullane}, {Grebel}, {Holland}, {Huc},
  {Passot}, {Bramante}, {Cacciari}, {Casta{\~n}eda}, {Chaoul}, {Cheek}, {De
  Angeli}, {Fabricius}, {Guerra}, {Hern{\'a}ndez}, {Jean-Antoine-Piccolo},
  {Masana}, {Messineo}, {Mowlavi}, {Nienartowicz}, {Ord{\'o}{\~n}ez-Blanco},
  {Panuzzo}, {Portell}, {Richards}, {Riello}, {Seabroke}, {Tanga},
  {Th{\'e}venin}, {Torra}, {Els}, {Gracia-Abril}, {Comoretto},
  {Garcia-Reinaldos}, {Lock}, {Mercier}, {Altmann}, {Andrae}, {Astraatmadja},
  {Bellas-Velidis}, {Benson}, {Berthier}, {Blomme}, {Busso}, {Carry},
  {Cellino}, {Clementini}, {Cowell}, {Creevey}, {Cuypers}, {Davidson}, {De
  Ridder}, {de Torres}, {Delchambre}, {Dell'Oro}, {Ducourant}, {Fr{\'e}mat},
  {Garc{\'\i}a-Torres}, {Gosset}, {Halbwachs}, {Hambly}, {Harrison}, {Hauser},
  {Hestroffer}, {Hodgkin}, {Huckle}, {Hutton}, {Jasniewicz}, {Jordan},
  {Kontizas}, {Korn}, {Lanzafame}, {Manteiga}, {Moitinho}, {Muinonen},
  {Osinde}, {Pancino}, {Pauwels}, {Petit}, {Recio-Blanco}, {Robin}, {Sarro},
  {Siopis}, {Smith}, {Smith}, {Sozzetti}, {Thuillot}, {van Reeven}, {Viala},
  {Abbas}, {Abreu Aramburu}, {Accart}, {Aguado}, {Allan}, {Allasia},
  {Altavilla}, {{\'A}lvarez}, {Alves}, {Anderson}, {Andrei}, {Anglada Varela},
  {Antiche}, {Antoja}, {Ant{\'o}n}, {Arcay}, {Atzei}, {Ayache}, {Bach},
  {Baker}, {Balaguer-N{\'u}{\~n}ez}, {Barache}, {Barata}, {Barbier}, {Barblan},
  {Baroni}, {Barrado y Navascu{\'e}s}, {Barros}, {Barstow}, {Becciani},
  {Bellazzini}, {Bellei}, {Bello Garc{\'\i}a}, {Belokurov}, {Bendjoya},
  {Berihuete}, {Bianchi}, {Bienaym{\'e}}, {Billebaud}, {Blagorodnova},
  {Blanco-Cuaresma}, {Boch}, {Bombrun}, {Borrachero}, {Bouquillon}, {Bourda},
  {Bouy}, {Bragaglia}, {Breddels}, {Brouillet}, {Br{\"u}semeister},
  {Bucciarelli}, {Budnik}, {Burgess}, {Burgon}, {Burlacu}, {Busonero}, {Buzzi},
  {Caffau}, {Cambras}, {Campbell}, {Cancelliere}, {Cantat-Gaudin}, {Carlucci},
  {Carrasco}, {Castellani}, {Charlot}, {Charnas}, {Charvet}, {Chassat},
  {Chiavassa}, {Clotet}, {Cocozza}, {Collins}, {Collins}, {Costigan}, {Crifo},
  {Cross}, {Crosta}, {Crowley}, {Dafonte}, {Damerdji}, {Dapergolas}, {David},
  {David}, {De Cat}, {de Felice}, {de Laverny}, {De Luise}, {De March}, {de
  Martino}, {de Souza}, {Debosscher}, {del Pozo}, {Delbo}, {Delgado},
  {Delgado}, {di Marco}, {Di Matteo}, {Diakite}, {Distefano}, {Dolding}, {Dos
  Anjos}, {Drazinos}, {Dur{\'a}n}, {Dzigan}, {Ecale}, {Edvardsson}, {Enke},
  {Erdmann}, {Escolar}, {Espina}, {Evans}, {Eynard Bontemps}, {Fabre},
  {Fabrizio}, {Faigler}, {Falc{\~a}o}, {Farr{\`a}s Casas}, {Faye}, {Federici},
  {Fedorets}, {Fern{\'a}ndez-Hern{\'a}ndez}, {Fernique}, {Fienga}, {Figueras},
  {Filippi}, {Findeisen}, {Fonti}, {Fouesneau}, {Fraile}, {Fraser}, {Fuchs},
  {Furnell}, {Gai}, {Galleti}, {Galluccio}, {Garabato}, {Garc{\'\i}a-Sedano},
  {Gar{\'e}}, {Garofalo}, {Garralda}, {Gavras}, {Gerssen}, {Geyer}, {Gilmore},
  {Girona}, {Giuffrida}, {Gomes}, {Gonz{\'a}lez-Marcos},
  {Gonz{\'a}lez-N{\'u}{\~n}ez}, {Gonz{\'a}lez-Vidal}, {Granvik}, {Guerrier},
  {Guillout}, {Guiraud}, {G{\'u}rpide}, {Guti{\'e}rrez-S{\'a}nchez}, {Guy},
  {Haigron}, {Hatzidimitriou}, {Haywood}, {Heiter}, {Helmi}, {Hobbs},
  {Hofmann}, {Holl}, {Holland}, {Hunt}, {Hypki}, {Icardi}, {Irwin}, {Jevardat
  de Fombelle}, {Jofr{\'e}}, {Jonker}, {Jorissen}, {Julbe}, {Karampelas},
  {Kochoska}, {Kohley}, {Kolenberg}, {Kontizas}, {Koposov}, {Kordopatis},
  {Koubsky}, {Kowalczyk}, {Krone-Martins}, {Kudryashova}, {Kull}, {Bachchan},
  {Lacoste-Seris}, {Lanza}, {Lavigne}, {Le Poncin-Lafitte}, {Lebreton},
  {Lebzelter}, {Leccia}, {Leclerc}, {Lecoeur-Taibi}, {Lemaitre}, {Lenhardt},
  {Leroux}, {Liao}, {Licata}, {Lindstr{\o}m}, {Lister}, {Livanou}, {Lobel},
  {L{\"o}ffler}, {L{\'o}pez}, {Lopez-Lozano}, {Lorenz}, {Loureiro},
  {MacDonald}, {Magalh{\~a}es Fernandes}, {Managau}, {Mann}, {Mantelet},
  {Marchal}, {Marchant}, {Marconi}, {Marie}, {Marinoni}, {Marrese},
  {Marschalk{\'o}}, {Marshall}, {Mart{\'\i}n-Fleitas}, {Martino}, {Mary},
  {Matijevi{\v{c}}}, {Mazeh}, {McMillan}, {Messina}, {Mestre}, {Michalik},
  {Millar}, {Miranda}, {Molina}, {Molinaro}, {Molinaro}, {Moln{\'a}r},
  {Moniez}, {Montegriffo}, {Monteiro}, {Mor}, {Mora}, {Morbidelli}, {Morel},
  {Morgenthaler}, {Morley}, {Morris}, {Mulone}, {Muraveva}, {Musella},
  {Narbonne}, {Nelemans}, {Nicastro}, {Noval}, {Ord{\'e}novic},
  {Ordieres-Mer{\'e}}, {Osborne}, {Pagani}, {Pagano}, {Pailler}, {Palacin},
  {Palaversa}, {Parsons}, {Paulsen}, {Pecoraro}, {Pedrosa}, {Pentik{\"a}inen},
  {Pereira}, {Pichon}, {Piersimoni}, {Pineau}, {Plachy}, {Plum}, {Poujoulet},
  {Pr{\v{s}}a}, {Pulone}, {Ragaini}, {Rago}, {Rambaux}, {Ramos-Lerate},
  {Ranalli}, {Rauw}, {Read}, {Regibo}, {Renk}, {Reyl{\'e}}, {Ribeiro},
  {Rimoldini}, {Ripepi}, {Riva}, {Rixon}, {Roelens}, {Romero-G{\'o}mez},
  {Rowell}, {Royer}, {Rudolph}, {Ruiz-Dern}, {Sadowski}, {Sagrist{\`a}
  Sell{\'e}s}, {Sahlmann}, {Salgado}, {Salguero}, {Sarasso}, {Savietto},
  {Schnorhk}, {Schultheis}, {Sciacca}, {Segol}, {Segovia}, {Segransan},
  {Serpell}, {Shih}, {Smareglia}, {Smart}, {Smith}, {Solano}, {Solitro},
  {Sordo}, {Soria Nieto}, {Souchay}, {Spagna}, {Spoto}, {Stampa}, {Steele},
  {Steidelm{\"u}ller}, {Stephenson}, {Stoev}, {Suess}, {S{\"u}veges}, {Surdej},
  {Szabados}, {Szegedi-Elek}, {Tapiador}, {Taris}, {Tauran}, {Taylor},
  {Teixeira}, {Terrett}, {Tingley}, {Trager}, {Turon}, {Ulla}, {Utrilla},
  {Valentini}, {van Elteren}, {Van Hemelryck}, {van Leeuwen}, {Varadi},
  {Vecchiato}, {Veljanoski}, {Via}, {Vicente}, {Vogt}, {Voss}, {Votruba},
  {Voutsinas}, {Walmsley}, {Weiler}, {Weingrill}, {Werner}, {Wevers},
  {Whitehead}, {Wyrzykowski}, {Yoldas}, {{\v{Z}}erjal}, {Zucker}, {Zurbach},
  {Zwitter}, {Alecu}, {Allen}, {Allende Prieto}, {Amorim},
  {Anglada-Escud{\'e}}, {Arsenijevic}, {Azaz}, {Balm}, {Beck}, {Bernstein},
  {Bigot}, {Bijaoui}, {Blasco}, {Bonfigli}, {Bono}, {Boudreault}, {Bressan},
  {Brown}, {Brunet}, {Bunclark}, {Buonanno}, {Butkevich}, {Carret}, {Carrion},
  {Chemin}, {Ch{\'e}reau}, {Corcione}, {Darmigny}, {de Boer}, {de Teodoro}, {de
  Zeeuw}, {Delle Luche}, {Domingues}, {Dubath}, {Fodor}, {Fr{\'e}zouls},
  {Fries}, {Fustes}, {Fyfe}, {Gallardo}, {Gallegos}, {Gardiol}, {Gebran},
  {Gomboc}, {G{\'o}mez}, {Grux}, {Gueguen}, {Heyrovsky}, {Hoar}, {Iannicola},
  {Isasi Parache}, {Janotto}, {Joliet}, {Jonckheere}, {Keil}, {Kim},
  {Klagyivik}, {Klar}, {Knude}, {Kochukhov}, {Kolka}, {Kos}, {Kutka}, {Lainey},
  {LeBouquin}, {Liu}, {Loreggia}, {Makarov}, {Marseille}, {Martayan},
  {Martinez-Rubi}, {Massart}, {Meynadier}, {Mignot}, {Munari}, {Nguyen},
  {Nordlander}, {Ocvirk}, {O'Flaherty}, {Olias Sanz}, {Ortiz}, {Osorio},
  {Oszkiewicz}, {Ouzounis}, {Palmer}, {Park}, {Pasquato}, {Peltzer}, {Peralta},
  {P{\'e}turaud}, {Pieniluoma}, {Pigozzi}, {Poels}, {Prat}, {Prod'homme},
  {Raison}, {Rebordao}, {Risquez}, {Rocca-Volmerange}, {Rosen}, {Ruiz-Fuertes},
  {Russo}, {Sembay}, {Serraller Vizcaino}, {Short}, {Siebert}, {Silva},
  {Sinachopoulos}, {Slezak}, {Soffel}, {Sosnowska}, {Strai{\v{z}}ys}, {ter
  Linden}, {Terrell}, {Theil}, {Tiede}, {Troisi}, {Tsalmantza}, {Tur},
  {Vaccari}, {Vachier}, {Valles}, {Van Hamme}, {Veltz}, {Virtanen}, {Wallut},
  {Wichmann}, {Wilkinson}, {Ziaeepour}, \& {Zschocke}}]{2016A&A...595A...1G}
{Gaia Collaboration}, {Prusti}, T., {de Bruijne}, J.~H.~J., {et~al.} 2016,
  \aap, 595, A1, \dodoi{10.1051/0004-6361/201629272}

\bibitem[{{Gao} {et~al.}(2010){Gao}, {Theuns}, {Frenk}, {Jenkins}, {Helly},
  {Navarro}, {Springel}, \& {White}}]{2010MNRAS.403.1283G}
{Gao}, L., {Theuns}, T., {Frenk}, C.~S., {et~al.} 2010, \mnras, 403, 1283,
  \dodoi{10.1111/j.1365-2966.2009.16225.x}

\bibitem[{{Genel} {et~al.}(2014){Genel}, {Vogelsberger}, {Springel}, {Sijacki},
  {Nelson}, {Snyder}, {Rodriguez-Gomez}, {Torrey}, \&
  {Hernquist}}]{2014MNRAS.445..175G}
{Genel}, S., {Vogelsberger}, M., {Springel}, V., {et~al.} 2014, \mnras, 445,
  175, \dodoi{10.1093/mnras/stu1654}

\bibitem[{{Griffen} {et~al.}(2018){Griffen}, {Dooley}, {Ji}, {O'Shea},
  {G{\'o}mez}, \& {Frebel}}]{2018MNRAS.474..443G}
{Griffen}, B.~F., {Dooley}, G.~A., {Ji}, A.~P., {et~al.} 2018, \mnras, 474,
  443, \dodoi{10.1093/mnras/stx2749}

\bibitem[{{Han} {et~al.}(2022){Han}, {Conroy}, {Johnson}, {Speagle}, {Bonaca},
  {Chandra}, {Naidu}, {Ting}, {Woody}, \& {Zaritsky}}]{2022arXiv220804327H}
{Han}, J.~J., {Conroy}, C., {Johnson}, B.~D., {et~al.} 2022, arXiv e-prints,
  arXiv:2208.04327.
\newblock \doarXiv{2208.04327}

\bibitem[{{Haywood} {et~al.}(2018){Haywood}, {Di Matteo}, {Lehnert}, {Snaith},
  {Khoperskov}, \& {G{\'o}mez}}]{2018ApJ...863..113H}
{Haywood}, M., {Di Matteo}, P., {Lehnert}, M.~D., {et~al.} 2018, \apj, 863,
  113, \dodoi{10.3847/1538-4357/aad235}

\bibitem[{{Hunter}(2007)}]{2007CSE.....9...90H}
{Hunter}, J.~D. 2007, Computing in Science and Engineering, 9, 90,
  \dodoi{10.1109/MCSE.2007.55}

\bibitem[{{Ishigaki} {et~al.}(2021){Ishigaki}, {Hartwig}, {Tarumi}, {Leung},
  {Tominaga}, {Kobayashi}, {Magg}, {Simionescu}, \&
  {Nomoto}}]{2021MNRAS.506.5410I}
{Ishigaki}, M.~N., {Hartwig}, T., {Tarumi}, Y., {et~al.} 2021, \mnras, 506,
  5410, \dodoi{10.1093/mnras/stab1982}

\bibitem[{{Jean-Baptiste} {et~al.}(2017){Jean-Baptiste}, {Di Matteo},
  {Haywood}, {G{\'o}mez}, {Montuori}, {Combes}, \&
  {Semelin}}]{2017A&A...604A.106J}
{Jean-Baptiste}, I., {Di Matteo}, P., {Haywood}, M., {et~al.} 2017, \aap, 604,
  A106, \dodoi{10.1051/0004-6361/201629691}

\bibitem[{{Kisku} {et~al.}(2021){Kisku}, {Schiavon}, {Horta}, {Mason},
  {Mackereth}, {Hasselquist}, {Garc{\'\i}a-Hern{\'a}ndez}, {Bizyaev},
  {Brownstein}, {Lane}, {Minniti}, {Pan}, \&
  {Roman-Lopes}}]{2021MNRAS.504.1657K}
{Kisku}, S., {Schiavon}, R.~P., {Horta}, D., {et~al.} 2021, \mnras, 504, 1657,
  \dodoi{10.1093/mnras/stab525}

\bibitem[{{Mackereth} {et~al.}(2019){Mackereth}, {Schiavon}, {Pfeffer},
  {Hayes}, {Bovy}, {Anguiano}, {Allende Prieto}, {Hasselquist}, {Holtzman},
  {Johnson}, {Majewski}, {O'Connell}, {Shetrone}, {Tissera}, \&
  {Fern{\'a}ndez-Trincado}}]{2019MNRAS.482.3426M}
{Mackereth}, J.~T., {Schiavon}, R.~P., {Pfeffer}, J., {et~al.} 2019, \mnras,
  482, 3426, \dodoi{10.1093/mnras/sty2955}

\bibitem[{{Majewski} {et~al.}(2017){Majewski}, {Schiavon}, {Frinchaboy},
  {Allende Prieto}, {Barkhouser}, {Bizyaev}, {Blank}, {Brunner}, {Burton},
  {Carrera}, {Chojnowski}, {Cunha}, {Epstein}, {Fitzgerald}, {Garc{\'\i}a
  P{\'e}rez}, {Hearty}, {Henderson}, {Holtzman}, {Johnson}, {Lam}, {Lawler},
  {Maseman}, {M{\'e}sz{\'a}ros}, {Nelson}, {Nguyen}, {Nidever}, {Pinsonneault},
  {Shetrone}, {Smee}, {Smith}, {Stolberg}, {Skrutskie}, {Walker}, {Wilson},
  {Zasowski}, {Anders}, {Basu}, {Beland}, {Blanton}, {Bovy}, {Brownstein},
  {Carlberg}, {Chaplin}, {Chiappini}, {Eisenstein}, {Elsworth}, {Feuillet},
  {Fleming}, {Galbraith-Frew}, {Garc{\'\i}a}, {Garc{\'\i}a-Hern{\'a}ndez},
  {Gillespie}, {Girardi}, {Gunn}, {Hasselquist}, {Hayden}, {Hekker}, {Ivans},
  {Kinemuchi}, {Klaene}, {Mahadevan}, {Mathur}, {Mosser}, {Muna}, {Munn},
  {Nichol}, {O'Connell}, {Parejko}, {Robin}, {Rocha-Pinto}, {Schultheis},
  {Serenelli}, {Shane}, {Silva Aguirre}, {Sobeck}, {Thompson}, {Troup},
  {Weinberg}, \& {Zamora}}]{2017AJ....154...94M}
{Majewski}, S.~R., {Schiavon}, R.~P., {Frinchaboy}, P.~M., {et~al.} 2017, \aj,
  154, 94, \dodoi{10.3847/1538-3881/aa784d}

\bibitem[{{Matsuno} {et~al.}(2019){Matsuno}, {Aoki}, \&
  {Suda}}]{2019ApJ...874L..35M}
{Matsuno}, T., {Aoki}, W., \& {Suda}, T. 2019, \apjl, 874, L35,
  \dodoi{10.3847/2041-8213/ab0ec0}

\bibitem[{{Matteucci}(2021)}]{2021A&ARv..29....5M}
{Matteucci}, F. 2021, \aapr, 29, 5, \dodoi{10.1007/s00159-021-00133-8}

\bibitem[{{McCarthy} {et~al.}(2012){McCarthy}, {Font}, {Crain}, {Deason},
  {Schaye}, \& {Theuns}}]{2012MNRAS.420.2245M}
{McCarthy}, I.~G., {Font}, A.~S., {Crain}, R.~A., {et~al.} 2012, \mnras, 420,
  2245, \dodoi{10.1111/j.1365-2966.2011.20189.x}

\bibitem[{{Monachesi} {et~al.}(2016{\natexlab{a}}){Monachesi}, {Bell},
  {Radburn-Smith}, {Bailin}, {de Jong}, {Holwerda}, {Streich}, \&
  {Silverstein}}]{2016MNRAS.457.1419M}
{Monachesi}, A., {Bell}, E.~F., {Radburn-Smith}, D.~J., {et~al.}
  2016{\natexlab{a}}, \mnras, 457, 1419, \dodoi{10.1093/mnras/stv2987}

\bibitem[{{Monachesi} {et~al.}(2016{\natexlab{b}}){Monachesi}, {G{\'o}mez},
  {Grand}, {Kauffmann}, {Marinacci}, {Pakmor}, {Springel}, \&
  {Frenk}}]{2016MNRAS.459L..46M}
{Monachesi}, A., {G{\'o}mez}, F.~A., {Grand}, R. J.~J., {et~al.}
  2016{\natexlab{b}}, \mnras, 459, L46, \dodoi{10.1093/mnrasl/slw052}

\bibitem[{{Monachesi} {et~al.}(2019){Monachesi}, {G{\'o}mez}, {Grand},
  {Simpson}, {Kauffmann}, {Bustamante}, {Marinacci}, {Pakmor}, {Springel},
  {Frenk}, {White}, \& {Tissera}}]{2019MNRAS.485.2589M}
---. 2019, \mnras, 485, 2589, \dodoi{10.1093/mnras/stz538}

\bibitem[{{Naidu} {et~al.}(2020){Naidu}, {Conroy}, {Bonaca}, {Johnson}, {Ting},
  {Caldwell}, {Zaritsky}, \& {Cargile}}]{2020ApJ...901...48N}
{Naidu}, R.~P., {Conroy}, C., {Bonaca}, A., {et~al.} 2020, \apj, 901, 48,
  \dodoi{10.3847/1538-4357/abaef4}

\bibitem[{{Nelson} {et~al.}(2019{\natexlab{a}}){Nelson}, {Pillepich},
  {Springel}, {Pakmor}, {Weinberger}, {Genel}, {Torrey}, {Vogelsberger},
  {Marinacci}, \& {Hernquist}}]{2019MNRAS.490.3234N}
{Nelson}, D., {Pillepich}, A., {Springel}, V., {et~al.} 2019{\natexlab{a}},
  \mnras, 490, 3234, \dodoi{10.1093/mnras/stz2306}

\bibitem[{{Nelson} {et~al.}(2019{\natexlab{b}}){Nelson}, {Springel},
  {Pillepich}, {Rodriguez-Gomez}, {Torrey}, {Genel}, {Vogelsberger}, {Pakmor},
  {Marinacci}, {Weinberger}, {Kelley}, {Lovell}, {Diemer}, \&
  {Hernquist}}]{2019ComAC...6....2N}
{Nelson}, D., {Springel}, V., {Pillepich}, A., {et~al.} 2019{\natexlab{b}},
  Computational Astrophysics and Cosmology, 6, 2,
  \dodoi{10.1186/s40668-019-0028-x}

\bibitem[{{Oliphant}(2007)}]{2007CSE.....9c..10O}
{Oliphant}, T.~E. 2007, Computing in Science and Engineering, 9, 10,
  \dodoi{10.1109/MCSE.2007.58}

\bibitem[{{Pillepich} {et~al.}(2015){Pillepich}, {Madau}, \&
  {Mayer}}]{2015ApJ...799..184P}
{Pillepich}, A., {Madau}, P., \& {Mayer}, L. 2015, \apj, 799, 184,
  \dodoi{10.1088/0004-637X/799/2/184}

\bibitem[{{Pillepich} {et~al.}(2018){Pillepich}, {Springel}, {Nelson}, {Genel},
  {Naiman}, {Pakmor}, {Hernquist}, {Torrey}, {Vogelsberger}, {Weinberger}, \&
  {Marinacci}}]{2018MNRAS.473.4077P}
{Pillepich}, A., {Springel}, V., {Nelson}, D., {et~al.} 2018, \mnras, 473,
  4077, \dodoi{10.1093/mnras/stx2656}

\bibitem[{{Pillepich} {et~al.}(2019){Pillepich}, {Nelson}, {Springel},
  {Pakmor}, {Torrey}, {Weinberger}, {Vogelsberger}, {Marinacci}, {Genel}, {van
  der Wel}, \& {Hernquist}}]{2019MNRAS.490.3196P}
{Pillepich}, A., {Nelson}, D., {Springel}, V., {et~al.} 2019, \mnras, 490,
  3196, \dodoi{10.1093/mnras/stz2338}

\bibitem[{{Planck Collaboration} {et~al.}(2016){Planck Collaboration}, {Ade},
  {Aghanim}, {Arnaud}, {Ashdown}, {Aumont}, {Baccigalupi}, {Banday},
  {Barreiro}, {Bartlett}, {Bartolo}, {Battaner}, {Battye}, {Benabed},
  {Beno{\^\i}t}, {Benoit-L{\'e}vy}, {Bernard}, {Bersanelli}, {Bielewicz},
  {Bock}, {Bonaldi}, {Bonavera}, {Bond}, {Borrill}, {Bouchet}, {Boulanger},
  {Bucher}, {Burigana}, {Butler}, {Calabrese}, {Cardoso}, {Catalano},
  {Challinor}, {Chamballu}, {Chary}, {Chiang}, {Chluba}, {Christensen},
  {Church}, {Clements}, {Colombi}, {Colombo}, {Combet}, {Coulais}, {Crill},
  {Curto}, {Cuttaia}, {Danese}, {Davies}, {Davis}, {de Bernardis}, {de Rosa},
  {de Zotti}, {Delabrouille}, {D{\'e}sert}, {Di Valentino}, {Dickinson},
  {Diego}, {Dolag}, {Dole}, {Donzelli}, {Dor{\'e}}, {Douspis}, {Ducout},
  {Dunkley}, {Dupac}, {Efstathiou}, {Elsner}, {En{\ss}lin}, {Eriksen},
  {Farhang}, {Fergusson}, {Finelli}, {Forni}, {Frailis}, {Fraisse},
  {Franceschi}, {Frejsel}, {Galeotta}, {Galli}, {Ganga}, {Gauthier}, {Gerbino},
  {Ghosh}, {Giard}, {Giraud-H{\'e}raud}, {Giusarma}, {Gjerl{\o}w},
  {Gonz{\'a}lez-Nuevo}, {G{\'o}rski}, {Gratton}, {Gregorio}, {Gruppuso},
  {Gudmundsson}, {Hamann}, {Hansen}, {Hanson}, {Harrison}, {Helou},
  {Henrot-Versill{\'e}}, {Hern{\'a}ndez-Monteagudo}, {Herranz}, {Hildebrandt},
  {Hivon}, {Hobson}, {Holmes}, {Hornstrup}, {Hovest}, {Huang}, {Huffenberger},
  {Hurier}, {Jaffe}, {Jaffe}, {Jones}, {Juvela}, {Keih{\"a}nen}, {Keskitalo},
  {Kisner}, {Kneissl}, {Knoche}, {Knox}, {Kunz}, {Kurki-Suonio}, {Lagache},
  {L{\"a}hteenm{\"a}ki}, {Lamarre}, {Lasenby}, {Lattanzi}, {Lawrence}, {Leahy},
  {Leonardi}, {Lesgourgues}, {Levrier}, {Lewis}, {Liguori}, {Lilje},
  {Linden-V{\o}rnle}, {L{\'o}pez-Caniego}, {Lubin}, {Mac{\'\i}as-P{\'e}rez},
  {Maggio}, {Maino}, {Mandolesi}, {Mangilli}, {Marchini}, {Maris}, {Martin},
  {Martinelli}, {Mart{\'\i}nez-Gonz{\'a}lez}, {Masi}, {Matarrese}, {McGehee},
  {Meinhold}, {Melchiorri}, {Melin}, {Mendes}, {Mennella}, {Migliaccio},
  {Millea}, {Mitra}, {Miville-Desch{\^e}nes}, {Moneti}, {Montier}, {Morgante},
  {Mortlock}, {Moss}, {Munshi}, {Murphy}, {Naselsky}, {Nati}, {Natoli},
  {Netterfield}, {N{\o}rgaard-Nielsen}, {Noviello}, {Novikov}, {Novikov},
  {Oxborrow}, {Paci}, {Pagano}, {Pajot}, {Paladini}, {Paoletti}, {Partridge},
  {Pasian}, {Patanchon}, {Pearson}, {Perdereau}, {Perotto}, {Perrotta},
  {Pettorino}, {Piacentini}, {Piat}, {Pierpaoli}, {Pietrobon}, {Plaszczynski},
  {Pointecouteau}, {Polenta}, {Popa}, {Pratt}, {Pr{\'e}zeau}, {Prunet},
  {Puget}, {Rachen}, {Reach}, {Rebolo}, {Reinecke}, {Remazeilles}, {Renault},
  {Renzi}, {Ristorcelli}, {Rocha}, {Rosset}, {Rossetti}, {Roudier},
  {Rouill{\'e} d'Orfeuil}, {Rowan-Robinson}, {Rubi{\~n}o-Mart{\'\i}n},
  {Rusholme}, {Said}, {Salvatelli}, {Salvati}, {Sandri}, {Santos},
  {Savelainen}, {Savini}, {Scott}, {Seiffert}, {Serra}, {Shellard}, {Spencer},
  {Spinelli}, {Stolyarov}, {Stompor}, {Sudiwala}, {Sunyaev}, {Sutton},
  {Suur-Uski}, {Sygnet}, {Tauber}, {Terenzi}, {Toffolatti}, {Tomasi},
  {Tristram}, {Trombetti}, {Tucci}, {Tuovinen}, {T{\"u}rler}, {Umana},
  {Valenziano}, {Valiviita}, {Van Tent}, {Vielva}, {Villa}, {Wade}, {Wandelt},
  {Wehus}, {White}, {White}, {Wilkinson}, {Yvon}, {Zacchei}, \&
  {Zonca}}]{2016A&A...594A..13P}
{Planck Collaboration}, {Ade}, P.~A.~R., {Aghanim}, N., {et~al.} 2016, \aap,
  594, A13, \dodoi{10.1051/0004-6361/201525830}

\bibitem[{{Posti} \& {Helmi}(2019)}]{2019A&A...621A..56P}
{Posti}, L., \& {Helmi}, A. 2019, \aap, 621, A56,
  \dodoi{10.1051/0004-6361/201833355}

\bibitem[{{Purcell} {et~al.}(2010){Purcell}, {Bullock}, \&
  {Kazantzidis}}]{2010MNRAS.404.1711P}
{Purcell}, C.~W., {Bullock}, J.~S., \& {Kazantzidis}, S. 2010, \mnras, 404,
  1711, \dodoi{10.1111/j.1365-2966.2010.16429.x}

\bibitem[{{Qu} {et~al.}(2011){Qu}, {Di Matteo}, {Lehnert}, {van Driel}, \&
  {Jog}}]{2011A&A...535A...5Q}
{Qu}, Y., {Di Matteo}, P., {Lehnert}, M.~D., {van Driel}, W., \& {Jog}, C.~J.
  2011, \aap, 535, A5, \dodoi{10.1051/0004-6361/201116502}

\bibitem[{{Rodriguez-Gomez} {et~al.}(2015){Rodriguez-Gomez}, {Genel},
  {Vogelsberger}, {Sijacki}, {Pillepich}, {Sales}, {Torrey}, {Snyder},
  {Nelson}, {Springel}, {Ma}, \& {Hernquist}}]{2015MNRAS.449...49R}
{Rodriguez-Gomez}, V., {Genel}, S., {Vogelsberger}, M., {et~al.} 2015, \mnras,
  449, 49, \dodoi{10.1093/mnras/stv264}

\bibitem[{{Rodriguez-Gomez} {et~al.}(2016){Rodriguez-Gomez}, {Pillepich},
  {Sales}, {Genel}, {Vogelsberger}, {Zhu}, {Wellons}, {Nelson}, {Torrey},
  {Springel}, {Ma}, \& {Hernquist}}]{2016MNRAS.458.2371R}
{Rodriguez-Gomez}, V., {Pillepich}, A., {Sales}, L.~V., {et~al.} 2016, \mnras,
  458, 2371, \dodoi{10.1093/mnras/stw456}

\bibitem[{{Searle} \& {Zinn}(1978)}]{1978ApJ...225..357S}
{Searle}, L., \& {Zinn}, R. 1978, \apj, 225, 357, \dodoi{10.1086/156499}

\bibitem[{{Sestito} {et~al.}(2019){Sestito}, {Longeard}, {Martin},
  {Starkenburg}, {Fouesneau}, {Gonz{\'a}lez Hern{\'a}ndez}, {Arentsen},
  {Ibata}, {Aguado}, {Carlberg}, {Jablonka}, {Navarro}, {Tolstoy}, \&
  {Venn}}]{2019MNRAS.484.2166S}
{Sestito}, F., {Longeard}, N., {Martin}, N.~F., {et~al.} 2019, \mnras, 484,
  2166, \dodoi{10.1093/mnras/stz043}

\bibitem[{{Sijacki} {et~al.}(2015){Sijacki}, {Vogelsberger}, {Genel},
  {Springel}, {Torrey}, {Snyder}, {Nelson}, \&
  {Hernquist}}]{2015MNRAS.452..575S}
{Sijacki}, D., {Vogelsberger}, M., {Genel}, S., {et~al.} 2015, \mnras, 452,
  575, \dodoi{10.1093/mnras/stv1340}

\bibitem[{{Steinmetz} {et~al.}(2006){Steinmetz}, {Zwitter}, {Siebert},
  {Watson}, {Freeman}, {Munari}, {Campbell}, {Williams}, {Seabroke}, {Wyse},
  {Parker}, {Bienaym{\'e}}, {Roeser}, {Gibson}, {Gilmore}, {Grebel}, {Helmi},
  {Navarro}, {Burton}, {Cass}, {Dawe}, {Fiegert}, {Hartley}, {Russell},
  {Saunders}, {Enke}, {Bailin}, {Binney}, {Bland-Hawthorn}, {Boeche}, {Dehnen},
  {Eisenstein}, {Evans}, {Fiorucci}, {Fulbright}, {Gerhard}, {Jauregi}, {Kelz},
  {Mijovi{\'c}}, {Minchev}, {Parmentier}, {Pe{\~n}arrubia}, {Quillen}, {Read},
  {Ruchti}, {Scholz}, {Siviero}, {Smith}, {Sordo}, {Veltz}, {Vidrih}, {von
  Berlepsch}, {Boyle}, \& {Schilbach}}]{2006AJ....132.1645S}
{Steinmetz}, M., {Zwitter}, T., {Siebert}, A., {et~al.} 2006, \aj, 132, 1645,
  \dodoi{10.1086/506564}

\bibitem[{{Tissera} {et~al.}(2014){Tissera}, {Beers}, {Carollo}, \&
  {Scannapieco}}]{2014MNRAS.439.3128T}
{Tissera}, P.~B., {Beers}, T.~C., {Carollo}, D., \& {Scannapieco}, C. 2014,
  \mnras, 439, 3128, \dodoi{10.1093/mnras/stu181}

\bibitem[{{van der Walt} {et~al.}(2011){van der Walt}, {Colbert}, \&
  {Varoquaux}}]{2011CSE....13b..22V}
{van der Walt}, S., {Colbert}, S.~C., \& {Varoquaux}, G. 2011, Computing in
  Science and Engineering, 13, 22, \dodoi{10.1109/MCSE.2011.37}

\bibitem[{{Vogelsberger} {et~al.}(2014{\natexlab{a}}){Vogelsberger}, {Genel},
  {Springel}, {Torrey}, {Sijacki}, {Xu}, {Snyder}, {Nelson}, \&
  {Hernquist}}]{2014MNRAS.444.1518V}
{Vogelsberger}, M., {Genel}, S., {Springel}, V., {et~al.} 2014{\natexlab{a}},
  \mnras, 444, 1518, \dodoi{10.1093/mnras/stu1536}

\bibitem[{{Vogelsberger} {et~al.}(2014{\natexlab{b}}){Vogelsberger}, {Genel},
  {Springel}, {Torrey}, {Sijacki}, {Xu}, {Snyder}, {Bird}, {Nelson}, \&
  {Hernquist}}]{2014Natur.509..177V}
---. 2014{\natexlab{b}}, \nat, 509, 177, \dodoi{10.1038/nature13316}

\bibitem[{{Waskom} {et~al.}(2020){Waskom}, {Botvinnik}, {Ostblom}, {Lukauskas},
  {Hobson}, {MaozGelbart}, {Gemperline}, {Augspurger}, {Halchenko}, {Cole},
  {Warmenhoven}, {De Ruiter}, {Pye}, {Hoyer}, {Vanderplas}, {Villalba},
  {Kunter}, {Quintero}, {Bachant}, {Martin}, {Meyer}, {Swain}, {Miles},
  {Brunner}, {O'Kane}, {Yarkoni}, {Williams}, \& {Evans}}]{2020zndo...3629446W}
{Waskom}, M., {Botvinnik}, O., {Ostblom}, J., {et~al.} 2020, {mwaskom/seaborn:
  v0.10.0 (January 2020)}, v0.10.0,  Zenodo, \dodoi{10.5281/zenodo.3629446}

\bibitem[{{Waters} {et~al.}(2024){Waters}, {Peterson}, {Emami}, {Shen},
  {Hernquist}, {Smith}, {Vogelsberger}, {Alcock}, {Tremblay}, {Liska},
  {Forbes}, \& {Moreno}}]{2024ApJ...961..193W}
{Waters}, T.~K., {Peterson}, C., {Emami}, R., {et~al.} 2024, \apj, 961, 193,
  \dodoi{10.3847/1538-4357/ad165a}

\bibitem[{{Weinberger} {et~al.}(2017){Weinberger}, {Springel}, {Hernquist},
  {Pillepich}, {Marinacci}, {Pakmor}, {Nelson}, {Genel}, {Vogelsberger},
  {Naiman}, \& {Torrey}}]{2017MNRAS.465.3291W}
{Weinberger}, R., {Springel}, V., {Hernquist}, L., {et~al.} 2017, \mnras, 465,
  3291, \dodoi{10.1093/mnras/stw2944}

\bibitem[{{Yanny} {et~al.}(2009){Yanny}, {Rockosi}, {Newberg}, {Knapp},
  {Adelman-McCarthy}, {Alcorn}, {Allam}, {Allende Prieto}, {An}, {Anderson},
  {Anderson}, {Bailer-Jones}, {Bastian}, {Beers}, {Bell}, {Belokurov},
  {Bizyaev}, {Blythe}, {Bochanski}, {Boroski}, {Brinchmann}, {Brinkmann},
  {Brewington}, {Carey}, {Cudworth}, {Evans}, {Evans}, {Gates}, {G{\"a}nsicke},
  {Gillespie}, {Gilmore}, {Nebot Gomez-Moran}, {Grebel}, {Greenwell}, {Gunn},
  {Jordan}, {Jordan}, {Harding}, {Harris}, {Hendry}, {Holder}, {Ivans},
  {Ivezi{\v{c}}}, {Jester}, {Johnson}, {Kent}, {Kleinman}, {Kniazev},
  {Krzesinski}, {Kron}, {Kuropatkin}, {Lebedeva}, {Lee}, {French Leger},
  {L{\'e}pine}, {Levine}, {Lin}, {Long}, {Loomis}, {Lupton}, {Malanushenko},
  {Malanushenko}, {Margon}, {Martinez-Delgado}, {McGehee}, {Monet}, {Morrison},
  {Munn}, {Neilsen}, {Nitta}, {Norris}, {Oravetz}, {Owen}, {Padmanabhan},
  {Pan}, {Peterson}, {Pier}, {Platson}, {Re Fiorentin}, {Richards}, {Rix},
  {Schlegel}, {Schneider}, {Schreiber}, {Schwope}, {Sibley}, {Simmons},
  {Snedden}, {Allyn Smith}, {Stark}, {Stauffer}, {Steinmetz}, {Stoughton},
  {SubbaRao}, {Szalay}, {Szkody}, {Thakar}, {Sivarani}, {Tucker}, {Uomoto},
  {Vanden Berk}, {Vidrih}, {Wadadekar}, {Watters}, {Wilhelm}, {Wyse}, {Yarger},
  \& {Zucker}}]{2009AJ....137.4377Y}
{Yanny}, B., {Rockosi}, C., {Newberg}, H.~J., {et~al.} 2009, \aj, 137, 4377,
  \dodoi{10.1088/0004-6256/137/5/4377}

\bibitem[{{Zolotov} {et~al.}(2009){Zolotov}, {Willman}, {Brooks}, {Governato},
  {Brook}, {Hogg}, {Quinn}, \& {Stinson}}]{2009ApJ...702.1058Z}
{Zolotov}, A., {Willman}, B., {Brooks}, A.~M., {et~al.} 2009, \apj, 702, 1058,
  \dodoi{10.1088/0004-637X/702/2/1058}

\bibitem[{{Zolotov} {et~al.}(2010){Zolotov}, {Willman}, {Brooks}, {Governato},
  {Hogg}, {Shen}, \& {Wadsley}}]{2010ApJ...721..738Z}
---. 2010, \apj, 721, 738, \dodoi{10.1088/0004-637X/721/1/738}

\end{thebibliography}
\bibliographystyle{aasjournal}

\end{document}